\documentclass[sigconf, nonacm, pdfa]{acmart}

\renewcommand{\paragraph}[1]{\vspace{0.3em}
\noindent\textbf{#1.}}
\renewcommand{\subparagraph}[1]{\vspace{0.3em}
\noindent\textit{\underline{#1.}}}

\usepackage{enumitem}
\usepackage[a-2b]{pdfx}

\setlist[itemize]{leftmargin=1.5em}
\setlist[enumerate]{leftmargin=1.5em}
\usepackage{booktabs}
\usepackage{marginnote}

\newcommand{\oursystem}[1]
{NeurIDA}

\usepackage{xspace}
\newcommand{\dime}{DIME\xspace}

\newcommand{\term}[1]{{\color{black}#1}}

\newlist{todolist}{itemize}{2}
\setlist[todolist]{label=$\square$}

\usepackage{pifont}
\usepackage{booktabs}
\usepackage{multirow}
\usepackage{graphicx}
\usepackage{subfig}

\usepackage[normalem]{ulem}

\usepackage{tikz}
\newcommand{\myarrow}[1][]{%
  \begin{tikzpicture}[#1]%
    \draw (0,0.7ex) -- (0,0) -- (0.75em,0);
    \draw (0.55em,0.2em) -- (0.75em,0) -- (0.55em,-0.2em);
  \end{tikzpicture}%
}

\newcommand{\ignore}[1]{}
\usepackage[flushleft]{threeparttable}

\usepackage{wrapfig}
\usepackage{listings}
\usepackage[most]{tcolorbox}

\usepackage{multicol}

\newtcolorbox[auto counter]{mybox}[1][]{%
breakable,
enhanced,
sharp corners,
colback=white,
fonttitle=\bfseries,
enlarge bottom at break by=5mm,
enlarge top at break by=5mm,
overlay first={%
    \draw[black, line width=0.5mm](frame.south west)--(frame.south east);
    \node[anchor=north east] at (frame.south east) {continued on next page};
    },
overlay middle={%
    \draw[black, line width=0.5mm](frame.south west)--(frame.south east);
    \draw[black, line width=0.5mm](frame.north west)--(frame.north east);
    \node[anchor=north east] at (frame.south east) {continued on next page};
    \node[anchor=south west] at (frame.north west) {continued from next page};
    },
overlay last={%
    \draw[black, line width=0.5mm](frame.north west)--(frame.north east);
    \node[anchor=south west] at (frame.north west) {continued from next page};},
#1
}

  {\list{}{\leftmargin=0.15in\rightmargin=0.15in}\item[]}%
  {\endlist}

\usepackage{multicol}

\definecolor{metacolor}{HTML}{000000} 
\definecolor{R1color}{HTML}{000000} 
\definecolor{R2color}{HTML}{000000} 
\definecolor{R3color}{HTML}{000000} 
\newcommand{\revone}[1]{\textcolor{R1color}{#1}}
\newcommand{\revtwo}[1]{\textcolor{R2color}{#1}}
\newcommand{\revthree}[1]{\textcolor{R3color}{#1}}

\newtcolorbox{myquoteMeta}[1][]{
    colback=black!5,
    colframe=black!5,
    notitle,
    sharp corners,
    borderline west={1.5pt}{0pt}{blue},
    enhanced,
    breakable,
    left=2pt,
    right=2pt,
    top=0pt,
    bottom=0pt,
    ignore nobreak,
}

\newtcolorbox{myquoteR1}[1][]{
    colback=black!5,
    colframe=black!5,
    notitle,
    sharp corners,
    borderline west={1.5pt}{0pt}{R1color},
    enhanced,
    breakable,
    left=2pt,
    right=2pt,
    top=0pt,
    bottom=0pt,
    ignore nobreak,
}

\newtcolorbox{myquoteR2}[1][]{
    colback=black!5,
    colframe=black!5,
    notitle,
    sharp corners,
    borderline west={1.5pt}{0pt}{R2color},
    enhanced,
    breakable,
    left=2pt,
    right=2pt,
    top=0pt,
    bottom=0pt,
    ignore nobreak,
    #1
}

\newtcolorbox{myquoteR3}[1][]{
    colback=black!3,
    colframe=black!3,
    notitle,
    sharp corners,
    borderline west={1.5pt}{0pt}{R3color},
    enhanced,
    breakable,
    left=2pt,
    right=2pt,
    top=0pt,
    bottom=0pt,
    ignore nobreak,
}

\begin{document}

\title{\oursystem{}: Dynamic Modeling for Effective In-Database Analytics}

\author{Lingze Zeng}
\email{lingze@comp.nus.edu.sg}
\affiliation{%
  \institution{National University of Singapore}
  \city{}
  \state{}
  \country{}
}

\author{Shaofeng Cai}
\email{shaofeng@comp.nus.edu.sg}
\affiliation{%
  \institution{National University of Singapore}
  \city{}
  \state{}
  \country{}
}

\author{Naili Xing}
\authornote{Corresponding author.}
\email{xingnl@comp.nus.edu.sg}
\affiliation{%
 \institution{National University of Singapore}
 \city{}
  \state{}
 \country{}
}

\author{Jiaqi Zhu}
\email{jiaqi77@nus.edu.sg}
\affiliation{%
  \institution{National University of Singapore}
  \city{}
  \state{}
  \country{}
}

\author{Gang Chen}
\email{cg@zju.edu.cn}
\affiliation{%
  \institution{Zhejiang University}
  \city{}
  \state{}
  \country{}
}

\author{Peng Lu}
\email{peng.lu@zju.edu.cn}
\affiliation{%
  \institution{Zhejiang University}
  \city{}
  \state{}
  \country{}
}

\author{Jian Pei}
\email{j.pei@duke.edu}
\affiliation{%
  \institution{Duke University}
  \city{}
  \state{}
  \country{}
}

\author{Beng Chin Ooi}
\email{ooibc@zju.edu.cn}
\affiliation{%
  \institution{Zhejiang University}
  \city{}
  \state{}
  \country{}
}

\begin{abstract}

Relational Database Management Systems (RDBMS) manage complex, interrelated data and support a broad spectrum of analytical tasks.
With the growing demand for predictive analytics, the deep integration of machine learning (ML) into RDBMS has become critical.
However, a fundamental challenge hinders this evolution: conventional ML models are static and task-specific, whereas RDBMS environments are dynamic and must support diverse analytical queries.
Each analytical task entails constructing a bespoke pipeline from scratch, which incurs significant development overhead and hence limits the wide adoption of ML in analytics.

We present \oursystem{}, an autonomous end-to-end system for in-database analytics that dynamically ``tweaks'' the best available base model to better serve a given analytical task.
In particular, we propose a novel paradigm of \textit{dynamic in-database modeling} to pre-train a composable base model architecture over the relational data.
Upon receiving a task, \oursystem{} formulates the task and data profile to dynamically select and configure relevant components from the pool of base models and shared model components for prediction.
For a friendly user experience, 
\oursystem{} supports natural language queries; it interprets user intent to construct structured task profiles and generates analytical reports with dedicated LLM agents. 
By design, \oursystem{} enables ease-of-use and yet effective and efficient in-database AI analytics.
Extensive experimental studies show that \oursystem{} consistently delivers up to 12\% improvement in AUC-ROC and 25\% relative reduction in MAE across ten tasks on five real-world datasets. 


\end{abstract}

\maketitle




\begin{figure*}[t] 
\centering 
\includegraphics[width=0.99\linewidth]{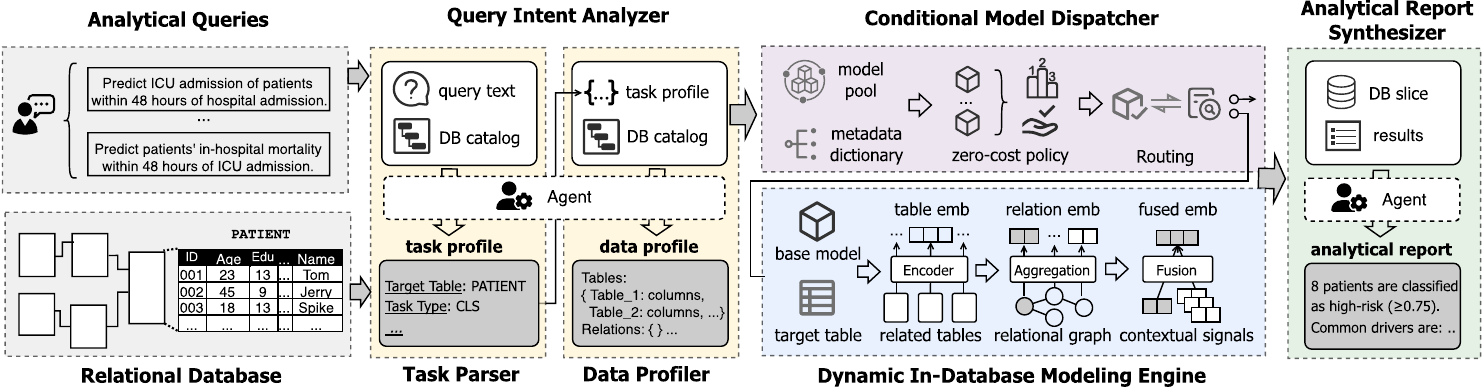}
\caption{The workflow and key components of \oursystem{}. 
}
\label{fig:overview}
\label{fig:workflow}
\end{figure*}

\section{Introduction}\label{sec:introduction}

Relational Database Management Systems (RDBMS) form the cornerstone of modern data analytics, enabling complex analytical workflows via structured query language (SQL)~\cite{SelingerACLP79,StonebrakerR86,Chaudhuri98,MoerkotteN08}.
The deep integration of machine learning (ML) represents the next frontier that promises to transform these systems from passive data repositories into proactive analytical engines~\cite{Brodie88,raven, moko, flowdb}.
However, while data-centric systems increasingly embed ML models within RDBMS, a fundamental paradigm gap between the static nature of ML and the dynamic environment of RDBMS has hindered this evolution, preventing the full realization of in-database AI-driven analytics~\cite{zhao2024cedar,pan2025database}.

The root of this challenge lies in the inherent divide in their designs~\cite{neurdb}.
RDBMS are built for dynamism, designed to support a diverse and evolving range of analytical tasks on continuously updating data~\cite{AkenPGZ17,KurmanjiTT24}.
In contrast, conventional ML models are inherently rigid, typically trained for a single, predefined prediction task on a static dataset snapshot and remain fixed after deployment~\cite{WuI24,KurmanjiT23}.
Consequently, when a new analytical task is issued, an existing model often cannot be readily adapted.
For instance, in an e-commerce database, a model built to predict customer churn from user behavior is ill-suited to forecasting product returns based on item reviews~\cite{de2025explainability}.
This model-level inflexibility necessitates the costly and inefficient practice of building a bespoke data-to-model pipeline from scratch for each new analytical requirement.

Current approaches to bridge this gap typically fall into two main strategies.
(1) Retraining a general-purpose model for each new task: This approach is resource-intensive and it undermines integrated analytics by requiring manual, external data processing that disrupts the in-database workflow~\cite{azure,postgresml,mindsdb}.
(2) Maintaining a pool of pre-trained models and selecting the "best fit" at runtime: This approach is constrained by the inherent inflexibility of the candidate models and the non-trivial challenge of curating a relevant pool amid continuously evolving data~\cite{Singa2015, trails,Cerebro,zhu2025context,zhu2026catching}.
Clearly, both strategies are fundamentally limited by their reliance on static models.
They lack the capacity for 
fine-grained ``dynamic'' customization
based on the specific semantics of an analytical query, such as its accessed tables, filter predicates, and therefore fail to consistently deliver optimal performance across diverse tasks~\cite{leads}.

Overcoming these limitations demands a new paradigm for ML-RDBMS integration that moves beyond static, embedded ML models towards a truly dynamic architecture.
This requires a system capable of
(1) dynamically constructing models tailored to individual analytical tasks,
(2) operating directly over native relational data to eliminate complex preprocessing,
and (3) providing a unified interface for seamless task formulation and result interpretation.
Such a system would ultimately render analytical workflows autonomous, efficient, and user-friendly.

In this paper, we present \oursystem{}, a \textbf{\uline{Neur}}\-al \textbf{\uline{I}}n-\textbf{\uline{D}}\-at\-a\-base \textbf{\uline{A}}\-nal\-y\-tics system that realizes these requirements through a novel paradigm of \textit{dynamic in-database modeling}.
To autonomously handle diverse analytical tasks end-to-end, as illustrated in Figure~\ref{fig:workflow}, the workflow of \oursystem{} is orchestrated by four key components:
First, given a \textsf{natural language query} (NLQ), the Query Intent Analyzer provides the unified interface that parses user intent to formulate the task, producing a structured task and data profile.
The Conditional Model Dispatcher then selects the best available \textsf{base model} and conditionally invokes dynamic modeling for model augmentation.
Next, the \term{Dynamic In-database Modeling Engine (\dime)} operates on the relational database to retrieve the required data and execute the model prediction. It either directly deploys the selected base model for efficiency or dynamically constructs a bespoke model to augment the base model for better serving the given task.
Finally, the Analytical Report Synthesizer handles result interpretation, synthesizing the prediction results into a comprehensive technical report delivered to the user.

\revtwo{At the core of \oursystem{} is \dime, which shifts modeling from a static design to a composable paradigm.}
When model augmentation is invoked, \dime dynamically constructs a bespoke model at query time using the selected \textsf{base model} and shared modules.
This dynamic modeling is a multi-stage process explicitly guided by the task:
First, \dime generates \textsf{tuple embeddings} via \textit{base table embedding} to capture individual-table semantics for the retrieved data.
Next, it constructs a task-specific \term{\textsf{relational graph}} to incorporate the inter-table structure via \textit{dynamic relation modeling}, producing enriched \term{relation embeddings}.
Finally, it integrates these embeddings via \textit{dynamic model fusion} into a unified \term{fused embedding} for the final \textit{task-specific prediction}.
This modeling process is conditioned on the query, where the task profile determines the model's structural composition, and the data profile provides the context to parametrically adjust computation, thereby establishing a flexible \term{\textsf{base model architecture}} capable of adaptively augmenting the base model for various analytical tasks.

In summary, we make the following contributions:
\begin{itemize}

\item \revtwo{We present \oursystem{}, an autonomous system for end-to-end in-database analytics within RDBMS, which enables the automatic execution of analytical tasks without domain expertise.}

\item We establish a composable model architecture, a modular framework that dynamically constructs bespoke models at query time by assembling base models and shared modules, as guided by the task and data profile.

\item We introduce the  dynamic in-database modeling paradigm that shifts from static models
to runtime model construction, reconciling the rigidity of ML with the dynamism of RDBMS.

\item 
We evaluate \oursystem{} on five real-world relational databases across ten analytical tasks. Extensive experiments show that \oursystem{} achieves up to 12\% improvement in AUC-ROC on classification tasks and a 10\%–25\% 
reduction in MAE on regression tasks 
with respect
to standalone base models.
\end{itemize}

\oursystem{} is a component of NeurDB~\cite{ooi2024neurdb,neurdb}.
It has been integrated into NeurDB as its native in-database analytics engine.

The remainder of this paper is structured as follows: Section~\ref{sec:preliminaries} introduces preliminaries. 
Section~\ref{sec:design} presents the design of \oursystem{}, detailing its four key modules and the dynamic model construction technique.
Section~\ref{sec:experiments} reports experimental results, and Section~\ref{sec:related_work} reviews related work.
Finally, Section~\ref{sec:conclusion} concludes the paper.

\section{Preliminaries}
\label{sec:preliminaries}

This section formally defines the foundational concepts for dynamic in-database modeling, It first details the structure of relational data and its schema, and then introduces the notion of a data slice.

\subsubsection*{\textbf{Relational Data}}
A relational database $\mathcal{D}$ is a structured collection of data organized into $K$ tables, denoted as $\mathcal{D}:= \{\mathcal{T}_k\}_{k=1}^K$.
Each table $\mathcal{T}_k$ represents a specific entity type (e.g., users, products) and consists of a set of $N_k$ rows (tuples) and $M_k$ columns (attributes).
A row $\mathcal{T}_{k,i:}$ represents a single data instance, denoted as the vector $\mathbf{x}_i = (x_1, x_2, \dots, x_{M^k})$ by assuming a canonical attribute ordering.
These attribute values can be of heterogeneous types, e.g., numerical, categorical, text, or timestamp.
A column $\mathcal{T}_{k,:j}$ corresponds to a specific attribute, holding values of a homogeneous data type for all tuples in the table.
Tables are interconnected through Primary Key (PK) and Foreign Key (FK) constraints, which enforce relational integrity and specify the structural dependencies among tables.

\subsubsection*{\textbf{Data Slice}}
Given an analytical query $q$, the \term{\textsf{data slice}} $\mathcal{D}_q$ is defined as the task-specific subset of the relational database relevant to this query, effectively forming a smaller but self-contained database derived from $\mathcal{D}$.
For instance, in a user-churn prediction task, the data slice would consist of only the tables and attributes related to user profiles and past purchases.
Formally, a data slice is a collection of \term{\textsf{table slices}}, defined as $\mathcal{D}_q = \{\mathcal{T}_{k,q} | \mathcal{T}_{k,q} \in \mathcal{D}, k\in \mathcal{K}_q \subseteq \{1,\cdots, K\}\}$,
where a table slice $\mathcal{T}_{k,q}$ is derived from its corresponding source table $\mathcal{T}_{k} \in \mathcal{D}$ by selecting a subset of its rows and columns.
This operation can be expressed as: $\mathcal{T}_{k,q} = \pi_{J_{k,q}}(\sigma_{I_{k,q}}(\mathcal{T}_k))$, where $\sigma$ is the selection operator for rows with indices $I_{k,q}$, and $\pi$ is the projection operator for columns with indices $J_{k,q}$, with both index sets determined by the SQL generated from query $q$.
The data slice $\mathcal{D}_q$ thus serves as the central object for our analytics pipeline, with all subsequent modeling performed exclusively on this focused subset.

\section{\oursystem{} Overview}
\label{sec:design}





\oursystem{} is designed to extend the capabilities of a traditional RDBMS to support dynamic in-database analytics.
It accepts 
a \textsf{natural language query} (NLQ) as input 
and automatically generates a comprehensive 
analytical
report as output.
The workflow of 
\oursystem{}, as illustrated in Figure~\ref{fig:overview}, 
is orchestrated by four key components:
(1) First, the \textbf{Query Intent Analyzer} parses the NLQ to formalize the \term{\textsf{task profile}} and the \term{\textsf{data profile}}, which specify the prediction task and required \textsf{data slice}, respectively.
(2) Next, the \textbf{Conditional Model Dispatcher} evaluates the task complexity to select the best \textsf{base model}, and decides whether to directly deploy the base model or invoke advanced model augmentation.
(3) Subsequently, \textbf{Dynamic In-Database Modeling Engine} (\dime)
retrieves the \textsf{data slice}, and executes the modeling based on the decision, either directly running the base model or dynamically constructing a bespoke model for the task.
(4) Finally, the \textbf{Analytical Report Synthesizer} synthesizes the prediction results within the context of the task to produce an analytical report.
We shall detail each component in subsequent sections.

\subsection{Query Intent Analyzer}
\label{sec:analyzer}
\revone{
Upon the arrival of an NLQ, the \textbf{Task Parser} and \textbf{Data Profiler} sequentially generate two structured outputs: a \emph{task profile} and a \emph{data profile}. 
The former defines the analytical task for subsequent modeling, while the latter specifies the data slice required for analytics.
}

\subsubsection*{\textbf{Task Parser}} 
\revone{
The Task Parser takes the NLQ and the \textsf{database catalog} (DB catalog) as input, which describes the schema, including tables and relationships. 
It formulates the \emph{task profile} implied in the query by identifying four key fields: the target table, task type, prediction window, and target query. 
Specifically, the target table denotes the primary entity table inferred from the NLQ and grounded in the database schema. The task type specifies the prediction objective, such as classification or regression. 
The prediction window defines the temporal horizon of the task, while the target query extracts the labeled data from the database for model training and evaluation. 
In \oursystem{}, this extraction process is automated by an analytical agent powered by a Large Language Model (LLM)~\cite{abs-2302-13971} using a carefully crafted prompt, as illustrated in Figure~\ref{fig:parser}. 
The prompt provides structured guidance by incorporating both the NLQ and the DB catalog as context.
To ensure output consistency, the prompt includes illustrative examples demonstrating the desired response format. The agent's response is parsed into the task profile and subsequently validated by a rule checker to guarantee syntactic correctness and schema alignment.
}

\subsubsection*{\textbf{Data Profiler}}
\begin{figure}[t] 
\centering 
\includegraphics[width=0.495\textwidth]{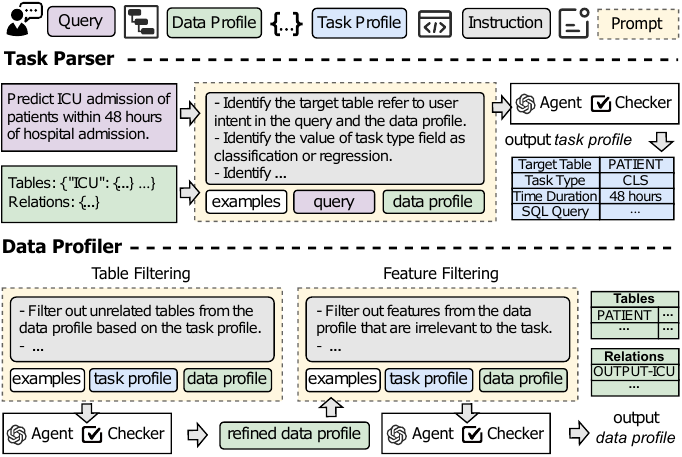}
\caption{\revone{Workflow of the Query Intent Analyzer, illustrating how an NLQ is parsed into structured task and data profiles.}}
\label{fig:parser}
\end{figure}

\revone{
Given the \textsf{task profile}, the Data Profiler constructs a \textsf{data profile} for subsequent modeling, which identifies the task-relevant subset of tables, relationships, and columns. In essence, the Data Profiler acts as a domain expert that curates the optimal schema context for the predictive task.
To automate this complex reasoning, \oursystem{} again employs an LLM-based analytical agent.
A chain-of-thought (CoT) prompting strategy is adopted to decompose the profiling task into multiple 
rounds of interactions, as illustrated at the bottom of Figure~\ref{fig:parser}.
The agent first identifies the related tables connected to the target table defined in the \emph{task profile}. 
Then, it refines the selected tables by filtering out irrelevant columns, such as internal database metadata or attributes lacking predictive signal.
%
The resulting data profile is integrated with the task profile to drive the subsequent modeling operations. 
}


\subsection{Conditional Model Dispatcher}
\label{sec:dispatcher}
Before invoking the \dime engine, \oursystem{} employs a lightweight \textbf{Conditional Model Dispatcher} to optimize efficiency for data analytics.
This component addresses two problems:
(1) Base Model Selection, determining which pre-trained base model $m_i$ from the model pool $\mathcal{M} = \{m_1, m_2, \dots, m_K\}$ is best suited for the current analytical task;
and (2) Conditional Augmentation, deciding whether the base model $m_i$ provides sufficient performance, or requires the relational augmentation via \dime, i.e., transforming $m_i$ into a dynamically constructed relational model, as invoking model augmentation could be unnecessary for simple tasks where a standalone base model can excel.
Therefore, the dispatcher maintains a metadata dictionary $\mathcal{D}_{p}$ that tracks the historical performance statistics, particularly the Exponential Moving Average (EMA) of each \textsf{base model} $m_i$, denoted as $\mu_i$.
\revone{
Upon receiving the \emph{task profile} and the \emph{data profile}, the dispatcher first constructs a small evaluation batch by randomly sampling labeled tuples from the data slice.
It then employs Zero-Cost Proxies (ZCP)~\cite{zcp, nasi}, a training-free evaluation technique originally proposed in Neural Architecture Search (NAS)~\cite{understandnas, anytimenas}, to efficiently estimate the predictive performance of each candidate base model without any training.
ZCPs estimate model quality at initialization by analyzing gradient and activation statistics over the sampled evaluation batch. In practice, each candidate model performs a single forward pass and an optional backward pass to compute a proxy score $s_i$, which correlates with the model's expected full-training performance.
The candidates are subsequently ranked by their proxy scores, and the one with the highest score $s^*$, is selected as the base model, $m^*$.
}

The dispatcher then determines whether to invoke the augmentation by comparing $s^*$ against a dynamic threshold $\tau$, 
\revone{
defined as $\tau = (1-\epsilon) \cdot \mu_{m^*}$, where  $\mu_{m^*}$ is the historical EMA of $m^*$ on the task and $\epsilon$ is a tolerance parameter controlling acceptable performance deviation.
}
If $s^* \ge \tau$, indicating that the base model maintains its performance within an acceptable margin of its historical standard, the system bypasses structural augmentation and directly deploys $m^*$ for prediction.
Otherwise, if $s^* < \tau$, the system invokes \dime to augment $m^*$ with relational context, dynamically constructing a bespoke model (denoted as $m^*$ w/ \oursystem{}) to bridge the performance gap.
\revone{This component ensures that \oursystem{} judiciously allocates computational resources, triggering augmentation only when the performance drop exceeds the tolerance threshold $\epsilon$.}

\begin{table}[t]
\caption{\revone{Notations used in the Conditional Model Dispatcher.}}
\label{tab:dispatcher_notation}
\renewcommand{\arraystretch}{0.92}
\setlength{\tabcolsep}{4pt}
\begin{tabular}{|c|p{0.75\columnwidth}|}
\hline
Notation & \multicolumn{1}{c|}{Description} \\ \hline
$m_i$ & Candidate base model. \\
$\mu_i$ & Historical EMA performance of $m_i$. \\
$s_i$ & Zero-Cost Proxy (ZCP) score of $m_i$. \\
$m^*$ & Optimal base model selected via ZCP. \\
$\tau$ & Selection threshold derived from the EMA of $m^*$. \\
$\epsilon$ & Performance tolerance parameter, $\epsilon\in[0,1]$. \\ \hline
\end{tabular}
\end{table}


\subsection{Dynamic In-database Modeling Engine}
\label{sec:dime_overview}
\begin{figure*}[t] 
\centering 
\includegraphics[width=0.92\textwidth]{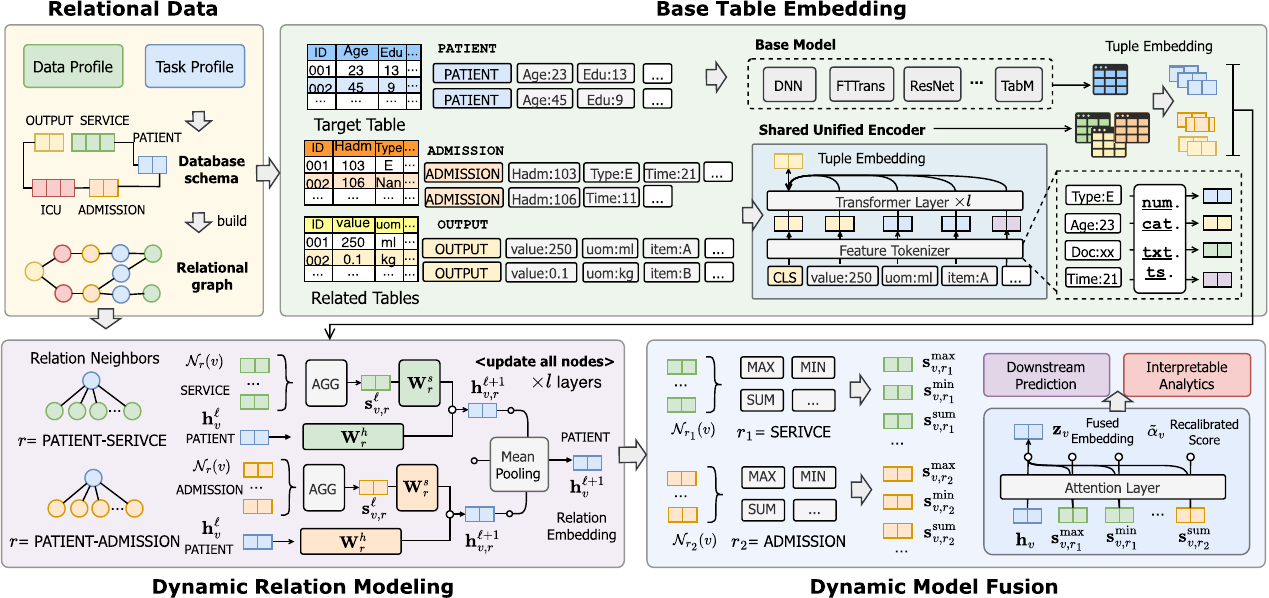}
\caption{Overview of 
the Dynamic In-Database Modeling Engine.}
\label{fig:modeling}
\end{figure*}

Once the modeling strategy is determined, \dime retrieves the corresponding \textsf{data slice} and dynamically constructs a bespoke model for the analytical task.
A key challenge in such dynamic modeling over relational databases is the extreme variability of analytical tasks and their associated data profiles. 
Static modeling is inherently limited: a single general-purpose model is inevitably suboptimal across diverse tasks, whereas training a dedicated model from scratch for each new task is prohibitively expensive.
To resolve this, \dime introduces a dynamic modeling paradigm.
At its core is a modular and composable framework that comprises a pool of heterogeneous \textsf{base models} and shared \textsf{model components}.
Given the task and data profile, \dime dynamically instantiates the predictive model by augmenting the selected base model with shared components.
\revthree{
This framework establishes a composable modeling backbone that jointly captures individual-table semantics and inter-table relational structures.
By design, this architecture enables adaptive modeling while maintaining transferability across diverse analytical tasks.}
The detailed execution flow of \dime is illustrated in Section~\ref{sec:dime}.


\subsection{Analytical Report Synthesizer}
\label{sec:synthesizer}
\oursystem{} generates predictions for tuples in the target table via \dime.
However, these outputs remain at an instance-level and require further interpretation to support high-level decision-making.
Translating these results into actionable insights, such as the drivers of customer churn or emerging market trends, traditionally requires analysis and summarization by domain experts.


To automate this process,  this component formulates the problem as an evidence-grounded generation task. Given the \emph{task profile}, the \emph{data slice}, and the prediction results, an analytical agent powered by LLMs examines these results and retrieves supporting evidence from data, such as representative tuples, statistical patterns, or aggregated trends. Based on this evidence, the agent reasons about the underlying signals and composes a structured analytical report aligned with the user's NLQ.

\section{The \dime Modeling Framework}
\label{sec:dime}
In this section, we present the \dime modeling framework and its execution flow. When model augmentation is invoked, \dime dynamically constructs a task-specific model within the composable framework. It first represents the \textsf{data slice} as a \term{\textsf{relational graph}}, where tuples from the \term{\textsf{target table}} and \term{\textsf{related tables}} are connected through Primary-Foreign Key (PK-FK) relationships.
Based on this graph, \dime instantiates a bespoke model using the selected \term{\textsf{base model}} together with shared \term{{model components}}, and applies it to generate predictions for target tuples.

As illustrated in Figure~\ref{fig:modeling}, the modeling pipeline consists of four stages.
(1) \textbf{Base Table Embedding}: generates initial \textsf{tuple embeddings} to capture individual-table semantics for tuples in the \textsf{target table} and \textsf{related tables} via a selected \textsf{base model} and a \textit{unified tuple encoder}, respectively.
(2) \textbf{Dynamic Relation Modeling}: enriches the \textsf{tuple embedding} with inter-table structures from the \textsf{relational graph}, producing \textsf{relation embeddings} through a shared \term{\textit{relation-aware message passing}} module. 
(3) \textbf{Dynamic Model Fusion}: enhances the representation of each target tuple into a \textsf{fused embedding} by incorporating relational context from neighboring \textsf{related tables} through a \term{\textit{context-aware fusion}} module.
(4) \textbf{Task-Aware Prediction}: feeds the representation of the target tuple into a task-specific prediction head for the final prediction.
\revtwo{
In the following subsections, we describe each component and explain how a bespoke model is dynamically constructed from these model components across different databases and tasks.}

\subsubsection*{\textbf{Base Table Embedding}}
The primary objective of this stage is to capture \term{\textit{intra-table semantics}} for all tuples in the retrieved \textsf{data slice}, providing the initial representation required for the subsequent relational augmentation.
Given the selected \textsf{base model} $m^*$, \dime employs a dual-path embedding strategy.
First, $m^*$ is applied to its original input features to produce native representations that preserve its inherent inductive biases.
In parallel, \dime introduces a shared model component, the \textit{unified tuple encoder}, which generates representations for all tuples across heterogeneous tables. The output vectors from both the base model and the unified tuple encoder are collectively referred to as \textsf{tuple embeddings} $\mathbf{e}$.
%
Specifically, the unified tuple encoder transforms tuples from respective table slices into embeddings in a common $d$-dimensional representation space.
Given a tuple $\mathbf{x} = (x_1, x_2, \cdots, x_M)$ with $M$ attributes in a table slice $\mathcal{T}$, the encoder first encodes each raw attribute value $x_i$ into a dense embedding vector $\mathbf{e}_i$ corresponding to its data type.
Specifically, for categorical attributes, we use a learnable embedding lookup table, i.e., $\mathbf{e}_i = \mathbf{E}_i[x_i]$. 
For numerical attributes, we apply a linear transformation with learnable parameters, i.e., $\mathbf{e}_i = x_i \cdot \hat{\mathbf{e}}_i + \mathbf{b}_i \quad \hat{\mathbf{e}}_i$ where $\mathbf{b}_i \in \mathbb{R}^d$. 
For textual attributes, we concatenate the column name with the attribute value, encode the resulting text using a shared pre-trained language model $\text{LM}(\cdot)$, and then project it to dimension $d$ via a linear layer, i.e., $\mathbf{e}_i = \text{Linear}(\text{LM}([{\tt col\_name}, x_i]))$.
For timestamp attributes, we use a \textsf{Time2Vec}~\cite{posencoding} encoder to capture absolute progression and periodic patterns. Concretely, each timestamp is decomposed into an absolute part $t_{\text{year}}$ and a cyclical part $\mathbf{t}_{\text{cyc}} = [t_{\text{month}}, t_{\text{day}}, t_{\text{hour}}]$, and encoded as $\mathbf{e}_t = [\phi_{\text{abs}}(t_{\text{year}} - t_{\text{base}}), \phi_{\text{cyc}}(\mathbf{t}_{\text{cyc}})] \mathbf{W}_t + \mathbf{b}_t$, where $\phi_{\text{abs}}$ is a learnable scaling function, $\phi_{\text{cyc}}$ uses sinusoidal mappings for periodicity.
%
In this way, all heterogeneous attributes are projected into a unified latent space $\mathbb{R}^d$.

Next, the set of attribute embeddings $\{\mathbf{e}_1, \dots, \mathbf{e}_M\}$ is then processed by a transformer-based module consisting of a  Multi-Head Self-Attention (MSA) layer and a Feed-Forward Network (FFN). 
The self-attention mechanism models attribute interactions to capture higher-order dependencies, and its permutation-invariance naturally aligns with the unordered nature of attributes in a relational tuple.
A learnable table-specific \texttt{[CLS]} token $\mathbf{e}_{\text{cls}}$ is prepended for summarization.
Let $\mathbf{E} = [\mathbf{e}_{\text{cls}}, \mathbf{e}_1, \dots, \mathbf{e}_M] \in \mathbb{R}^{(M+1)\times d}$ denote the input sequence.
For attention head $h \in \{1, \dots, H\}$:
\begin{align*}
    \text{Attn}_{h}(\textbf{E}) = \text{softmax}(\frac{(\textbf{E}\textbf{W}_Q^{(h)})(\textbf{E}\textbf{W}_K^{(h)})^{\textbf{T}}}{\sqrt{d_h}}) (\textbf{E}\textbf{W}_V^{(h)})
\end{align*}
\noindent
where $\mathbf{W}_Q^{(h)}, \mathbf{W}_K^{(h)}, \mathbf{W}_V^{(h)} \in \mathbb{R}^{d\times d_h}$ are projection matrices.
The outputs are concatenated and projected back to dimension $d$:
\begin{align*}
    \text{MSA}(\textbf{E}) = \text{Concat} (\text{Attn}_1(\textbf{E}),\cdots, \text{Attn}_H(\textbf{E})) \textbf{W}_O
\end{align*}
\noindent
where $\mathbf{W}_O \in \mathbb{R}^{H \cdot d_h \times d}$ is the output projection.
Residual connections and layer normalization are applied to stabilize the embedding process:
$\textbf{H}' = \text{LayerNorm}(\textbf{E} + \text{MSA}(\textbf{E}))$.
The embeddings are then refined by a subsequent FFN layer: $\textbf{H} = \text{LayerNorm}(\textbf{H}' + \text{FFN}(\textbf{H}'))$.
Finally, the tuple embedding $\mathbf{c}$ is obtained by mean-pooling the output sequence $\mathbf{H}$, followed by layer normalization: $$\textbf{c} = \text{LayerNorm}(\text{MEAN}(\textbf{H}))$$
This representation captures fine-grained intra-table information through aggregated attribute interactions learned by the encoder.
By sharing this model component across diverse table slices, \dime establishes a highly scalable and unified interface for augmenting the base model in subsequent relational modeling.

Crucially, this design facilitates the seamless augmentation of any selected \textsf{base model}, e.g., FT-Transformer~\cite{revisiting} or ARM-Net~\cite{CaiZ0JOZ21}. By generating tuple embeddings via the unified tuple encoder for the supporting table slices, \dime effectively ``wraps'' the base model with rich inter-table structural context in subsequent stages.
The mechanism dynamically extends the capabilities of the base model, enabling it to exploit the full relational schema without altering its internal feature processing logic.

\subsubsection*{\textbf{Dynamic Relation Modeling}}
In this stage, the given \textsf{tuple embeddings} are refined by integrating \term{\textit{inter-table structures}}. Specifically, the task-specific \textsf{relational graph} is explicitly modeled to capture structural interactions within tuples defined by the underlying schema constraints.


Formally, the data slice is represented as a heterogeneous graph $\mathcal{G}=(\mathcal{V}, \mathcal{E}, \mathcal{R}, \mathcal{A})$, where $\mathcal{A}$ denotes node types (tables, e.g., \texttt{USER}, \texttt{ITEM}), and $\mathcal{R}$ represents the set of relation types (PK-FK links, e.g., \texttt{USER–PURCHASE–ITEM}).
Each node $v\in\mathcal{V}$ corresponds to a tuple, and each edge $(u,r,v) \in \mathcal{E}$ represents an undirected PK-FK link of relation type $r \in \mathcal{R}$.
The relation-specific neighborhood $\mathcal{N}_r(v)$ is defined as the set of nodes $u$ connected to $v$ via the relation type $r$: $\mathcal{N}_r(v) = \{ u | (u,r,v) \in \mathcal{E}\}$.
For each node $v$, the representation at layer $\ell = 0$ is initialized as $\mathbf{h}_{v}^{(\ell)} = \textbf{c}_{v}$ (the tuple embedding from the previous stage). 
A shared \term{\textit{relation-aware message passing}} module then propagates information along the relational graph. For each relation type $r$, neighbor messages are first aggregated:
\begin{align*}
    \mathbf{s}_{v,r}^{(\ell)} = \text{AGG}({\mathbf{h}_u^{(\ell)}: u\in\mathcal{N}_r(v)})
\end{align*}

\noindent
where $\text{AGG}(\cdot)$ is an aggregate function (e.g., mean or sum) that produces a relation-specific local summary $\mathbf{s}_{v,r}^{(\ell)}$. It is then integrated into the current node representation:
\begin{align*}
    \mathbf{h}_{v,r}^{(\ell + 1)} = \mathbf{W}_{r}^h\mathbf{h}_{v,r}^{(\ell)} +\mathbf{W}_{r}^s\mathbf{s}_{v,r}^{(\ell)}
\end{align*}
\noindent
where $\mathbf{W}_r^h,\mathbf{W}_r^s \in \mathbb{R}^{d \times d}$ are relation-specific learnable parameters.
In this design, the model captures different semantic roles across schema relations. For instance, \texttt{item–category} links reflect hierarchical structure, whereas \texttt{user–item} links represent interaction patterns. Maintaining relation-specific parameters enables the model to adaptively interpret neighbor information from highly diverse schema contexts.
%
Since a node may participate in multiple relations (e.g., a product linked to a category, an order, and a user review), its final representation $\mathbf{h}_{v}$ (\textsf{relational embedding}) consolidates messages across all relation types via a lightweight cross-relation aggregation with normalization and nonlinearity:
\begin{align*}
    \mathbf{h}_{v}^{(\ell + 1)} = \text{LayerNorm}(\phi(\frac{1}{|\mathcal{R}_{v}|}\sum_{r\in \mathcal{R}_{v}}\mathbf{h}_{v,r}^{(\ell+1)}))
\end{align*}
\noindent
where $\phi(\cdot)$ is the activation function, and $\mathcal{R}_{v}$ denotes the set of relation types associated with node $v$.
\revtwo{Notably, the resulting relational embedding remains instance-dependent: each node aggregates messages only from the active relation set and sampled context neighbors, reflecting the dynamically induced relational structure.
}

Overall, this shared model component enables tuple embeddings to incorporate rich structural information from the database schema.
By stacking multiple layers of this model component, the receptive field of each \textsf{relational embedding} naturally expands to encompass multi-hop neighborhoods. This allows the constructed model to encode higher-order interactions across the specified relational graph, thereby capturing inter-table dependencies beyond the immediate connections of individual tables.

\subsubsection*{\textbf{Dynamic Model Fusion}}
\label{sec:fusion}
While the preceding dynamic relation modeling stage implicitly enriches tuple representations through structural message passing, this stage explicitly synthesizes a task-specific summary from the relational neighborhood of the target tuple for final prediction.
At this point, the \textsf{relational embedding} $\mathbf{h}_{v}$ already incorporates both \term{intra-table semantic} and \term{inter-table structural information}.
However, maximizing predictive performance requires distinguishing which specific relational statistics, e.g., the \textit{maximum} purchase versus the \textit{average} rating in a repurchase task, are more discriminative for the task at hand.
%


To this end, we introduce a \emph{context-aware fusion} module, which adaptively fuses the target tuple embedding with distinct aggregated relational contexts. 
Specifically, for each \textsf{target tuple} $v$, a set of relation-specific contextual signals is generated via multiple aggregation functions to the neighbors $\mathcal{N}_r(v)$.

\begin{align*}
    \textbf{s}_{v, r}^{AGG} =  \text{AGG}(\{ \textbf{h}_u : u \in \mathcal{N}_r(v)\})
\end{align*}

\noindent
where $\text{AGG}(\cdot) \in \mathcal{F}$ denotes a set of adopted aggregate functions (Max, Min, Sum, Mean).
Each resulting signal $\textbf{s}_{v, r}^{\text{AGG}}$ captures a distinct statistical property of neighbors connected via relation type $r$.
These signals are organized into a contextual sequence $\mathcal{S}_v$, which concatenates the target embedding $\mathbf{h}_v$ with all derived \term{contextual signals} $\textbf{s}_{v, r}^{\text{AGG}}$ across the associated relation types $r \in \mathcal{R}_v$:

\begin{align*}
    \mathcal{S}_v = [\mathbf{h}_v, \mathbf{s}_{v, r_1}^{\text{max}}, \dots, \mathbf{s}_{v, r_1}^{\text{sum}}, \dots, \mathbf{s}_{v, r_{|\mathcal{R}_v|}}^{\text{max}}, \dots, \mathbf{s}_{v, r_{|\mathcal{R}_v|}}^{\text{sum}}]
\end{align*}

\noindent
This sequence is then processed by a task-specific Multi-Head Self-Attention layer to capture interactions among these contextual signals.
The final \textsf{fused embedding} $\mathbf{z}_v$ is obtained by retrieving the output vector corresponding to the target tuple embedding $\mathbf{h}_v$: $$\mathbf{z}_v = \text{MSA}(\mathcal{S}_v) [0] $$
\revtwo{This module complements the previous stage by acting as a fine-grained summarizer. 
It is trained end-to-end for the target task, enabling selective attention over informative relational signals. The resulting \textsf{fused embedding} provides a more robust and task-specific representation for final prediction.
}


\vspace{1.5mm}
\noindent
\textit{Interpretability.}
\label{sec:interpretability}
Beyond improving prediction performance, the self-attention mechanism inherently supports model interpretability.
The attention weights generated during fusion form a probability distribution over the contextual sequence $\mathcal{S}_v$, indicating the relative contribution to the final \textsf{fused embedding} from different aggregated relational statistics $\textbf{s}_{v, r}^{AGG}$, such as payment sums or average ratings. 
%
By analyzing these weights, \dime offers transparency into the dynamically constructed model, revealing which specific relational signals are more influential for the given prediction task.

However, raw attention weights are not directly comparable across tuples, since the length of the contextual sequence $|\mathcal{S}_v|$ varies with the number of associated relation types.
For instance, the same weight may indicate high importance in a long sequence but only moderate importance in a short one.
To address this, \dime computes a \term{recalibrated importance score} $\tilde{\alpha}_{v,i}$ by quantifying the deviation of the raw weight $\alpha_{v,i}$ from the uninformative uniform baseline $b = 1/|\mathcal{S}_v|$:
$$\tilde{\alpha}_{v,i} = \max\left(\frac{\alpha_{v,i} - b}{1 - b}, 0\right), i \in \{0, 1, \dots, |\mathcal{S}_v|-1\}. $$

\noindent
This normalization provides a length-independent metric, facilitating comparable interpretability of importance across diverse relational signals.

\begin{table*}[]
\centering
\caption{\revone{Statistics of databases and prediction tasks.}}
\vspace{-2mm}
\label{tab:data}
\resizebox{0.85\textwidth}{!}{%
\begin{threeparttable}

\begin{tabular}{@{}cccclllllcclll@{}}
\toprule[1.5pt]
\multirow{2}{*}{\textbf{Dataset}} & \multirow{2}{*}{\textbf{\#Table}} & \multirow{2}{*}{\textbf{\#Relation}} & \multirow{2}{*}{\textbf{\#Column}} & \multicolumn{1}{c}{\multirow{2}{*}{\textbf{Domain}}} & \multicolumn{1}{c}{\multirow{2}{*}{\textbf{Task}}} & \multicolumn{1}{c}{\multirow{2}{*}{\revone{\textbf{Type}}}} & \multicolumn{2}{c}{\multirow{2}{*}{\textbf{\begin{tabular}[c]{@{}c@{}}Target Table\\ \#Instance\end{tabular}}}} & \multicolumn{2}{c}{\textbf{\#Attribute}}    & \multicolumn{3}{c}{\textbf{\#Instance}}                                          \\ \cmidrule(l){10-14} 
                                  &                                   &                                      &                                    & \multicolumn{1}{c}{}                                 & \multicolumn{1}{c}{}                               & \multicolumn{1}{c}{}                      & \multicolumn{2}{c}{}                                                                                            & $\text{Target}^\ast$ & $\text{All}^\dagger$ & \multicolumn{1}{c}{Train} & \multicolumn{1}{c}{Valid} & \multicolumn{1}{c}{Test} \\ \midrule
\multirow{2}{*}{\textbf{Event}}   & \multirow{2}{*}{5}                & \multirow{2}{*}{7}                   & \multirow{2}{*}{117}               & \multicolumn{1}{l|}{\multirow{2}{*}{Sociality}}      & user-repeat                                        & \revone{CLS}                            & users                                                   & 37,143                                                & 8                    & 25                   & 3,842                     & 268                       & 246                      \\
                                  &                                   &                                      &                                    & \multicolumn{1}{l|}{}                                & user-attend                                        & \revone{REG}                                & users                                                   & 37,143                                                & 8                    & 19                   & 19,239                    & 2,013                     & 1,958                    \\ \midrule
\multirow{2}{*}{\textbf{Beer}}    & \multirow{2}{*}{9}                & \multirow{2}{*}{12}                  & \multirow{2}{*}{116}               & \multicolumn{1}{l|}{\multirow{2}{*}{E-commerce}}     & user-active                                        & \revone{CLS}                            & users                                                   & 89,662                                                & 11                   & 47                   & 16,656                    & 2,794                     & 3,558                    \\
                                  &                                   &                                      &                                    & \multicolumn{1}{l|}{}                                & beer-score                                           & \revone{REG}                                & beers                                                   & 719,164                                               & 21                   & 43                   & 45,922                    & 12,858                    & 7,218                    \\ \midrule
\multirow{2}{*}{\textbf{Trial}}   & \multirow{2}{*}{15}               & \multirow{2}{*}{15}                  & \multirow{2}{*}{77}                & \multicolumn{1}{l|}{\multirow{2}{*}{Healthcare}}     & study-result                                      & \revone{CLS}                            & studies                                                 & 249,730                                               & 27                   & 50                   & 11,994                    & 960                       & 825                      \\
                                  &                                   &                                      &                                    & \multicolumn{1}{l|}{}                                & site-success                                       & \revone{REG}                                & facilities                                              & 453,233                                               & 7                    & 14                   & 100,000                   & 19,740                    & 22,617                   \\ \midrule
\multirow{2}{*}{\textbf{Avito}}   & \multirow{2}{*}{8}                & \multirow{2}{*}{11}                  & \multirow{2}{*}{26}                & \multicolumn{1}{l|}{\multirow{2}{*}{Online Ads}}     & user-clicks                                        & \revone{CLS}                            & user\_info                                                & 98,250                                                & 6                    & 15                   & 59,454                    & 21,183                    & 47,996                   \\
                                  &                                   &                                      &                                    & \multicolumn{1}{l|}{}                                & ad-ctr                                             & \revone{REG}                                & ad\_info                                                 & 5,960,558                                             & 8                    & 18                   & 5,100                     & 1,766                     & 1,816                    \\ \midrule
\multirow{2}{*}{\textbf{HM}}      & \multirow{2}{*}{3}                & \multirow{2}{*}{2}                   & \multirow{2}{*}{37}                & \multicolumn{1}{l|}{\multirow{2}{*}{Retail}}         & user-churn                                         & \revone{CLS}                            & customer                                                & 1,371,980                                             & 8                    & 13                   & 100,000                   & 76,556                    & 74,575                   \\
                                  &                                   &                                      &                                    & \multicolumn{1}{l|}{}                                & item-sales                                         & \revone{REG}                                & article                                                 & 105,542                                               & 26                   & 32                   & 100,000                   & 100,000                   & 100,000                  \\ \bottomrule[1.5pt]
\end{tabular}

\begin{tablenotes}
            \item $\ast$ (\#Attribute) Target: denotes the number of attributes available within the target table's schema.
            \item $\dagger$ (\#Attribute) All: represents the total cumulative number of unique attributes across the entire relational database (the union of all table schemas).
\end{tablenotes}
\end{threeparttable}
}
\vspace{-2mm}
\end{table*}

\subsubsection*{\textbf{Task-Aware Prediction}}
Finally, the fused embedding $\mathbf{z}_v$ of the \textsf{target tuple}  passes through an activation and normalization layer, serving as the input for a task-specific prediction head $g_\theta(\cdot)$ for final prediction:
\begin{align*}
    \hat{y} = g_\theta(\text{LayerNorm}(\phi(\mathbf{z}_v))).
\end{align*}

\noindent
Here, the prediction head $g_\theta(\cdot)$ and the output $\hat{y}$ are determined by the \emph{task profile}.
For instance, for a classification task, $\hat{y}$ represents the predicted class probability, while for a regression task, it corresponds to a continuous numerical score.

\vspace{1.5mm}
\noindent
\textit{Optimization.}
\revthree{To balance model transferability and adaptability across diverse analytical tasks, \dime employs a hybrid optimization strategy.
}
Foundational modules, i.e., the \textit{unified tuple encoder} and the \textit{relation-aware message passing} module, are pre-trained and shared across tasks to capture universal database patterns.
In contrast, task-specific model components, including the \textit{context-aware fusion} module, the prediction head, and any integrated advanced \textit{base model}, e.g., FT-Transformer~\cite{revisiting} or ARM-Net~\cite{CaiZ0JOZ21}, are optimized or fine-tuned in a task-specific manner to maximize performance for the current analytical task.
For binary classification tasks, the optimization function is the binary cross-entropy loss:
\begin{align*}
    \mathcal{L}({\hat{y}, y} ) = -\frac{1}{N} \sum_{i=1}^N \{ y_i  {\rm log} \sigma(\hat{y_i}) + 
    (1 - y_i) {\rm log} (1 - \sigma(\hat{y_i})) \}
\end{align*}

\noindent
For regression tasks, the optimization function employs the L1 loss:
\begin{align*} 
\mathcal{L}({\hat{y}, y}) = \frac{1}{N}\sum_{i=1}^N |\hat{y}_i - y_i|
\end{align*}

\noindent
where $y$ denotes the ground truth label and $N$ is the number of training instances in the \emph{data slice}.

\section{Experiments}
\label{sec:experiments}

In this section, we evaluate \oursystem{} through experiments on multiple real-world relational databases and various prediction tasks. We further analyze its modules through ablation and sensitivity studies, examine system cost, and quantify the agent-based components.

\subsection{Experimental Setup} \label{sec.exp.exp_setup}
\subsubsection*{\textbf{Datasets and Tasks}}
We conduct studies on five relational databases from healthcare, sociality, and e-commerce domains.
Table~\ref{tab:data} provides the statistics of databases and associated prediction tasks.

\vspace{0.5mm}
\noindent $\bullet$ \textbf{\textsc{Event}}~\cite{event} 
is a recommendation database from a mobile social-planning app, containing users, invitations, and event information. 
There are two tasks:  (i) predicting whether a user will re-attend a joined event (\textsf{user-repeat}), and (ii) estimating how many invitations the user will respond yes to in the next week (\textsf{user-attend}).

\vspace{0.5mm}
\noindent $\bullet$ \textbf{\textsc{Beer}}~\cite{beer} is from a beer-review platform that records users, beers, and reviews information. It involves two tasks: 
(i) predicting whether a user will write more than ten reviews in the next season (\textsf{user-active}), and (ii) predicting the positive-rating ratio, where reviews with scores above 3.5 are treated as positive (\textsf{beer-score}).

\vspace{0.5mm}
\noindent $\bullet$ \textbf{\textsc{Trial}}~\cite{trial}
is a clinical trial database
, containing studies, designs, and outcomes. We evaluate two tasks for the next year: (i) predicting whether a clinical study will be successful (\textsf{study-result}), and (ii) estimating the site-level success rate (\textsf{site-success}).

\vspace{0.5mm}
\noindent $\bullet$ \textbf{\textsc{Avito}}~\cite{avito} 
is a database of an online advertising platform that records user interactions such as searches, visits, and phone contacts. We consider two tasks for the next four days: (i) predicting whether a user will click on more than one advertisement (\textsf{user-clicks}), and (ii) estimating the click-through rate (\textsf{ad-ctr}).

\vspace{0.5mm}
\noindent $\bullet$ \textbf{\textsc{HM}}~\cite{hm} 
is a retail database containing customer profiles, item information, and transaction records. We define two tasks for the next week: (i) predicting whether a customer will churn, i.e., make no transaction (\textsf{user-churn}), and (ii) predicting the item-level sales volume, measured by total transaction value (\textsf{item-sales}).

\begin{table*}[]
\centering

\caption{Overall effectiveness of \oursystem{} across five relational databases.}
\label{tab:result_effect}
\resizebox{0.915\textwidth}{!}{%
\begin{threeparttable}
\begin{tabular}{@{}clcccccccccc@{}}
\toprule[1.5pt]
\multicolumn{2}{c}{Task Type (Metric)} &
  \multicolumn{5}{c}{Classification (AUC-ROC) $\uparrow$} &
  \multicolumn{5}{c}{Regression (MAE) $\downarrow$} \\ \midrule
\multicolumn{2}{c|}{Dataset /Task} &
  \multirow{2}{*}{\begin{tabular}[c]{@{}c@{}}Event\\ user-repeat\end{tabular}} &
  \multirow{2}{*}{\begin{tabular}[c]{@{}c@{}}Beer\\ user-active\end{tabular}} &
  \multirow{2}{*}{\begin{tabular}[c]{@{}c@{}}Trial\\ study-result\end{tabular}} &
  \multirow{2}{*}{\begin{tabular}[c]{@{}c@{}}Avito\\ 
  \textsf{user-clicks}\end{tabular}} &
  \multicolumn{1}{c|}{\multirow{2}{*}{\begin{tabular}[c]{@{}c@{}}HM\\ user-churn\end{tabular}}} &
  \multirow{2}{*}{\begin{tabular}[c]{@{}c@{}}Event\\ user-attend\end{tabular}} &
  \multirow{2}{*}{\begin{tabular}[c]{@{}c@{}}Beer\\ beer-score\end{tabular}} &
  \multirow{2}{*}{\begin{tabular}[c]{@{}c@{}}Trial\\ site-success\end{tabular}} &
  \multirow{2}{*}{\begin{tabular}[c]{@{}c@{}}Avito\\ ad-ctr\end{tabular}} &
  \multirow{2}{*}{\begin{tabular}[c]{@{}c@{}}HM\\ item-sales\end{tabular}} \\ \cmidrule(r){1-2}
Type &
  \multicolumn{1}{c|}{Method} &
   &
   &
   &
   &
  \multicolumn{1}{c|}{} &
   &
   &
   &
   &
   \\ \midrule
\multicolumn{1}{c|}{\multirow{6}{*}{STM}} &
  \multicolumn{1}{l|}{LR} &
  \multicolumn{1}{c}{0.7376} &
  \multicolumn{1}{c}{0.8812} &
  \multicolumn{1}{c}{0.6881} &
  \multicolumn{1}{c}{0.6407} &
  \multicolumn{1}{c|}{0.6182} &
  \multicolumn{1}{c}{0.3912} &
  \multicolumn{1}{c}{0.2046} &
  \multicolumn{1}{c}{0.4594} &
  \multicolumn{1}{c}{0.0483} &
  0.0659 \\
\multicolumn{1}{c|}{} &
  \multicolumn{1}{l|}{RF} &
  \multicolumn{1}{c}{0.7270} &
  \multicolumn{1}{c}{0.8737} &
  \multicolumn{1}{c}{0.6770} &
  \multicolumn{1}{c}{0.6450} &
  \multicolumn{1}{c|}{0.6104} &
  \multicolumn{1}{c}{0.3745} &
  \multicolumn{1}{c}{0.1997} &
  \multicolumn{1}{c}{0.4565} &
  \multicolumn{1}{c}{0.0485} &
  0.0584 \\
\multicolumn{1}{c|}{} &
  \multicolumn{1}{l|}{CatBoost} &
  \multicolumn{1}{c}{0.7429} &
  \multicolumn{1}{c}{0.9060} &
  \multicolumn{1}{c}{0.6945} &
  \multicolumn{1}{c}{0.6504} &
  \multicolumn{1}{c|}{0.6232} &
  \multicolumn{1}{c}{0.2607} &
  \multicolumn{1}{c}{0.1975} &
  \multicolumn{1}{c}{0.4429} &
  \multicolumn{1}{c}{0.0393} &
  0.0557 \\
\multicolumn{1}{c|}{} &
  \multicolumn{1}{l|}{LightGBM} &
  \multicolumn{1}{c}{0.7340} &
  \multicolumn{1}{c}{0.9061} &
  \multicolumn{1}{c}{0.6999} &
  \multicolumn{1}{c}{0.6527} &
  \multicolumn{1}{c|}{0.6241} &
  \multicolumn{1}{c}{0.2547} &
  \multicolumn{1}{c}{0.1903} &
  \multicolumn{1}{c}{0.4595} &
  \multicolumn{1}{c}{0.0379} &
  0.0495\\ \cmidrule(l){2-12} 
\multicolumn{1}{c|}{} &
\multicolumn{1}{l|}{TabPFN} & 
  0.7754 &
  0.9164 &
  0.7051 &
  0.6437 &
  \multicolumn{1}{c|}{0.6254} &
  0.3406 &
  0.1903 &
  0.4639 &
  0.0369 &
  0.0583 \\  
  \multicolumn{1}{c|}{} &
  \multicolumn{1}{l|}{TabICL\textsuperscript{$\ast$}} &
   0.7692&
   0.9077&
  0.7018&
  0.6503&
  \multicolumn{1}{c|}{0.6371} &
  -&
  -&
  -&
  -&
  -\\ 
  \midrule
\multicolumn{1}{c|}{\multirow{10}{*}{TRM}} &
  \multicolumn{1}{l|}{DNN} &
  0.7294 &
  0.9064 &
  0.6830 &
  0.6546 &
  \multicolumn{1}{c|}{0.6290} &
  0.2523 &
  0.1947 &
  0.4500 &
  0.0394 &
  0.0551 \\
\multicolumn{1}{c|}{} &
  \multicolumn{1}{l|}{\myarrow w/ \oursystem{}} &
  \textbf{0.7901} &
  \textbf{0.9289} &
 \textbf{0.7057}&
  \textbf{0.6714} &
  \multicolumn{1}{c|}{\textbf{0.6837}} &
 \textbf{0.2403} &
  \textbf{0.1719} &
  \textbf{0.3919} &
  \textbf{0.0366} &
  \textbf{0.0425} \\ \cmidrule(l){2-12} 
\multicolumn{1}{c|}{} &
  \multicolumn{1}{l|}{DeepFM} &
  0.7130 &
  0.9047 &
  0.7013&
  0.6283 &
  \multicolumn{1}{c|}{0.6203} &
  0.2491 &
  0.2091 &
  0.4581 &
  0.0397 &
  0.0541 \\
\multicolumn{1}{c|}{} &
  \multicolumn{1}{l|}{\myarrow w/ \oursystem{}} &
  \textbf{0.7993 }&
  \textbf{0.9366} &
  \textbf{0.7042} &
  \textbf{0.6737} &
  \multicolumn{1}{c|}{\textbf{0.6833}} &
 \textbf{0.2469} &
  \textbf{0.1727} &
  \textbf{0.4077 }&
 \textbf{0.0365} &
  \textbf{0.0419} \\ \cmidrule(l){2-12} 
\multicolumn{1}{c|}{} &
  \multicolumn{1}{l|}{ResNet} &
  0.7461 &
  0.9055 &
  0.7044 &
  0.6587 &
  \multicolumn{1}{c|}{0.6332} &
  0.2422 &
  0.1891 &
  0.4441 &
  0.0378 &
  0.0599 \\
\multicolumn{1}{c|}{} &
  \multicolumn{1}{l|}{\myarrow w/ \oursystem{}} &
  \textbf{0.7973 }&
  \textbf{0.9379} &
 \textbf{0.7097} &
  \textbf{0.6740} &
  \multicolumn{1}{c|}{\textbf{0.6874}} &
  \textbf{0.2376} &
 \textbf{0.1715} &
  \textbf{0.3713} &
  \textbf{0.0345} &
  \textbf{0.0390} \\ \cmidrule(l){2-12} 
\multicolumn{1}{c|}{} &
  \multicolumn{1}{l|}{FTTrans} &
  0.7346 &
  0.9131 &
  0.6836 &
  0.6502 &
  \multicolumn{1}{c|}{0.6304} &
  0.2539 &
  0.1825 &
  0.4290 &
  0.0374 &
  0.0584 \\
\multicolumn{1}{c|}{} &
  \multicolumn{1}{l|}{\myarrow w/ \oursystem{}} &
  \textbf{0.8012} &
  \textbf{0.9360} &
  \textbf{0.7171} &
  \textbf{0.6730} &
  \multicolumn{1}{c|}{\textbf{0.6655}} &
  \textbf{0.2401} &
  \textbf{0.1696} &
  \textbf{0.3817} &
  \textbf{0.0347} &
  \textbf{0.0398} \\ \cmidrule(l){2-12} 
\multicolumn{1}{c|}{} &
 \multicolumn{1}{l|}{ARM-Net} &
   0.7402&
  0.9016 &
  0.6965 &
  0.6604 &
  \multicolumn{1}{c|}{0.6297} &
  0.2642&
  0.1912&
  0.4468 &
  0.0368&
  0.0515\\
\multicolumn{1}{c|}{} &
  \multicolumn{1}{l|}{\myarrow w/ \oursystem{}} &
  \textbf{0.7934} &
  \textbf{0.9394} &
  \textbf{0.7130} &
  \textbf{0.6783} &
  \multicolumn{1}{c|}{\textbf{0.6900}} &
  \textbf{0.2372} &
  \textbf{0.1673} &
  \textbf{0.3906} &
  \textbf{0.0349} &
  \textbf{0.0403} \\ \cmidrule(l){2-12} 
\multicolumn{1}{c|}{} &
  \multicolumn{1}{l|}{TabM} &
  0.7415 &
  0.9014 &
  0.7048 &
  0.6610 &
  \multicolumn{1}{c|}{0.6270} &
  0.2505 &
  0.1829 &
  0.4087 &
  0.0371 &
  0.0510 \\
\multicolumn{1}{c|}{} &
  \multicolumn{1}{l|}{\myarrow w/ \oursystem{}} &
  \textbf{0.8028} &
  \textbf{0.9370} &
  \textbf{0.7166} &
  \textbf{0.6891 }&
  \multicolumn{1}{c|}{\textbf{0.6837}} &
  \textbf{0.2397} &
  \textbf{0.1692 }&
  \textbf{0.3812} &
  \textbf{0.0352} &
  \textbf{0.0408} \\ \midrule
\multicolumn{1}{c|}{\multirow{4}{*}{AutoML}} &
  \multicolumn{1}{l|}{Optuna} &
  0.7538 &
  0.9112 &
  0.7021 &
  0.6392 &
  \multicolumn{1}{c|}{0.6287} &
  0.2378 &
  0.1841 &
  0.4069 &
  0.0389 &
  0.0533 \\
\multicolumn{1}{c|}{} &
  \multicolumn{1}{l|}{\myarrow w/ \oursystem{}} &
  \textbf{0.7993} &
  \textbf{0.9317} &
  \textbf{0.7192} &
  \textbf{0.6755} &
\multicolumn{1}{c|}{\textbf{0.6793}} &
 \textbf{0.2335}&
  \textbf{0.1703}&
 \textbf{0.3891}&
  \textbf{0.0345}&
  \textbf{0.0401}\\ \cmidrule(l){2-12} 
\multicolumn{1}{c|}{} &
  \multicolumn{1}{l|}{Trials} &
  0.7602 &
  0.9130 &
  0.7040 &
  0.6568 &
  \multicolumn{1}{c|}{0.6225} &
  0.2480 &
  0.1919 &
  0.4023 &
  0.0401 &
  0.0497 \\
\multicolumn{1}{c|}{} &
  \multicolumn{1}{l|}{\myarrow w/ \oursystem{}} &
  \textbf{0.8017} &
 \textbf{0.9301} &
  \textbf{0.7142 }&
  \textbf{0.6657}&
  \multicolumn{1}{c|}{\textbf{0.6739}} &
  \textbf{0.2391} &
  \textbf{0.1710} &
  \textbf{0.3745} &
  \textbf{0.0373} &
  \textbf{0.0399} \\ \midrule
\multicolumn{1}{c|}{\multirow{6}{*}{LTM}} &
  \multicolumn{1}{l|}{TP-BERTa\textsuperscript{$\dagger$}} &
   0.5457&
   0.5170&
  -& 
  0.5122&
  \multicolumn{1}{c|}{0.5058} &
  0.2768&
  0.3155&
  0.4612&
  0.0430&
  0.3514\\
\multicolumn{1}{c|}{} &
  \multicolumn{1}{l|}{\myarrow w/ \oursystem{}} &
  \textbf{0.7705} &
 \textbf{0.9248} &
  \textbf{-}&
  \textbf{0.6472}&
  \multicolumn{1}{c|}{\textbf{0.6677}} &
  \textbf{0.2518} &
  \textbf{0.1864} &
  \textbf{0.4281} &
  \textbf{0.0403} &
  \textbf{0.0455} \\
    \cmidrule(l){2-12}
\multicolumn{1}{c|}{\multirow{3}{*}{}} &
  \multicolumn{1}{l|}{Nomic} &
  \multicolumn{1}{c}{0.6896} &
  \multicolumn{1}{c}{0.8896} &
  \multicolumn{1}{c}{0.6533} &
  \multicolumn{1}{c}{0.5771} &
  \multicolumn{1}{c|}{0.5313} &
  \multicolumn{1}{c}{0.2677} &
  \multicolumn{1}{c}{0.3439} &
  \multicolumn{1}{c}{0.4545} &
  \multicolumn{1}{c}{0.0435} &  
  0.2063 \\ 
\multicolumn{1}{c|}{} &
  \multicolumn{1}{l|}{\myarrow w/ \oursystem{}} &
  \multicolumn{1}{c}{\textbf{0.7730}} &
  \multicolumn{1}{c}{\textbf{0.9178}} &
  \multicolumn{1}{c}{\textbf{0.6760}} &
  \multicolumn{1}{c}{\textbf{0.6558}} &
  \multicolumn{1}{c|}{\textbf{0.6663}} &
  \multicolumn{1}{c}{\textbf{0.2504}} &
  \multicolumn{1}{c}{\textbf{0.1842}} &
  \multicolumn{1}{c}{\textbf{0.4228}} &
  \multicolumn{1}{c}{\textbf{0.0397}} &
  \textbf{0.0453} \\ 
  \cmidrule(l){2-12}
\multicolumn{1}{c|}{\multirow{3}{*}{}} &
  \multicolumn{1}{l|}{BGE} &
  \multicolumn{1}{c}{0.6787} &
  \multicolumn{1}{c}{0.8868} &
  \multicolumn{1}{c}{0.6503} &
  \multicolumn{1}{c}{0.6462} &
  \multicolumn{1}{c|}{0.6097} &
  \multicolumn{1}{c}{0.2645} &
  \multicolumn{1}{c}{0.2829} &
  \multicolumn{1}{c}{0.4511} &
  \multicolumn{1}{c}{0.0410} &
   0.0772 \\
\multicolumn{1}{c|}{} &
  \multicolumn{1}{c|}{\myarrow w/ \oursystem{}} &
  \multicolumn{1}{c}{\textbf{0.7857}} &
  \multicolumn{1}{c}{\textbf{0.9224}} &
  \multicolumn{1}{c}{\textbf{0.6805}} &
  \multicolumn{1}{c}{\textbf{0.6580}} &
  \multicolumn{1}{c|}{\textbf{0.6657}} &
  \multicolumn{1}{c}{\textbf{0.2474}} &
  \multicolumn{1}{c}{\textbf{0.1938}} &
  \multicolumn{1}{c}{\textbf{0.4098}} &
  \multicolumn{1}{c}{\textbf{0.0401}} &
  \textbf{0.0481} \\ 
  \bottomrule[1.5pt]
\end{tabular}%
\begin{tablenotes}
        \item $\ast$ TabICL: designed for classification, incompatible with regression tasks.
        \item $\dagger$ TP-BERTa: the sequence length required by the \textsf{Trial} \textsf{(study-result)} task exceeds the TP-BERTa maximum input limit (512 tokens).
\end{tablenotes}
\end{threeparttable}
}
\end{table*}

\subsubsection*{\textbf{Baselines}}
We use a variety of base models as baselines, which are organized into the following four groups:

\vspace{0.5mm}
\noindent $\bullet$ \textbf{Standalone Tabular Models (STM).}
This group operates on target tables and serves as a strong baseline for tabular analytics. 
It includes classical machine learning models such as Logistic Regression (\textsf{LR})~\cite{cox1958regression} and Random Forest (\textsf{RF})~\cite{Breiman01}, advanced tree-based models such as \textsf{CatBoost}~\cite{ProkhorenkovaGV18} and \textsf{LightGBM}~\cite{KeMFWCMYL17}, and recent tabular foundation models, including \textsf{TabPFN}~\cite{tabfpn} and \textsf{TabICL}~\cite{tabicl}, which adapt to new prediction tasks without training.

\vspace{0.5mm}
\noindent $\bullet$ \textbf{Tuple Representation Models (TRM).}
These neural tabular models learn tuple representations that can be further enhanced by \oursystem{} with relational structure. 
We include a diverse set of architectures in the model pool: \textsf{DNN}, \textsf{DeepFM}~\cite{GuoTYLH17}, \textsf{ResNet}~\cite{revisiting}, \textsf{FT-Transformer} (\textsf{FTTrans})~\cite{revisiting}, \textsf{ARM-Net}~\cite{CaiZ0JOZ21}, and \textsf{TabM}~\cite{GorishniyKB25}.

\vspace{0.5mm}
\noindent $\bullet$ \textbf{AutoML.}
To ensure competitive model configurations, we integrate automated learning approaches, including \textsf{Optuna}~\cite{AkibaSYOK19} for 
hyperparameter optimization and \textsf{Trails}~\cite{trails} for model architecture search. 
These approaches 
operate orthogonally to \oursystem{}.

\vspace{0.5mm}
\noindent $\bullet$ \textbf{Large Tabular Models (LTM).} 
We further evaluate pre-trained language models that treat each tuple as text and encode it into dense representations. LTM includes \textsf{TP-BERTa}~\cite{YanZXZC00C24}, a RoBERTa-based model for tabular analytics tasks, as well as \textsf{Nomic}~\cite{NussbaumMMD25} and \textsf{BGE}~\cite{XiaoLZMLN24}, two general text embedding models pre-trained on large-scale corpora for semantic representation learning.

\subsubsection*{\textbf{Evaluation}}
In the main study, we report \textit{the area under the curve} (\textsf{AUC-ROC}) for classification tasks, where a higher value indicates better discriminative performance. 
We use the \textit{mean absolute error} (\textsf{MAE}) for regression tasks, where a lower value denotes higher prediction accuracy.
To ensure temporal consistency and prevent information leakage, each dataset is chronologically partitioned into training, validation, and test sets based on global time cutoffs. The specific cutoff points vary across prediction tasks and databases.
%

To ensure fair and robust evaluation, all results are averaged over three independent runs.
We apply \textit{early stopping}~\cite{earlystop} on the validation set. Specifically, training stops if the validation metric does not improve for a predefined number of consecutive epochs, and the best-performing checkpoint is retained for testing.

\subsubsection*{\textbf{Implementation}}
\oursystem{} adopts a hybrid implementation strategy that balances in-database efficiency with LLM flexibility. 
Specifically, the performance-critical modeling components, namely the dispatcher and \dime, are implemented as user-defined functions (UDFs).
This native execution minimizes data movement and accelerates inference.
In contrast, the analyzer and synthesizer operate as external services to simplify LLM integration and management, as LLMs typically require substantial memory and GPU resources.

For the evaluation of \dime, all approaches are configured consistently. 
In particular, \textsf{FTTrans} uses 4 attention heads, \textsf{ARM-Net} uses 1 attention head, and \textsf{TabM} employs 8 ensemble experts to balance efficiency and accuracy.
For the AutoML approaches, \textsf{ResNet} serves as the backbone model, with the search space defined by the number of layers and hidden widths with candidate values in ${32, 64, 128, \ldots, 512}$. 
Both \textsf{Trails} and \textsf{Optuna} automatically explore the space to select the best architecture for each task.
%
For the LTM group, we treat each pre-trained model as a frozen encoder that maps a tuple into a semantic embedding. We use the public multitask checkpoint for \textsf{TP-BERTa}, and \textsf{nomic-embed-text-v1.5} and \textsf{bge-base-en-v1.5} for \textsf{Nomic} and \textsf{BGE}, respectively. 
All three models produce 768-dimensional embeddings, ensuring consistent representation sizes across LTM baselines.

In model training, we use the AdamW~\cite{adamw} with a learning rate searched within $[10^{-3}, 10^{-1}]$ and a batch size of 256. The patience epoch in \emph{early stopping} is set to  10  for regression and 5 for classification. 
%
All experiments are conducted on a server with a Xeon Silver 4114 CPU @ 2.2GHz (10 cores), 256GB of memory, and 8 GeForce RTX 3090 Ti. Model implementations are based on PyTorch 2.1.0~\cite{pytorch}, Pytorch Geometric 2.5.3~\cite{pyg}, with CUDA 11.8. 

\subsection{Overall Performance Evaluation} \label{sec.exp.main_res}
The main result is presented in Table~\ref{tab:result_effect}, where we evaluate the performance of base models across different prediction tasks. For models that support tuple representation learning, we further compare their performance with and without the augmentation of \oursystem{}. Based on these results, we draw the following key observations:

\begin{figure}[t]
\centering
\subfloat[\textsf{Avito (user-clicks)}]{
\includegraphics[width=0.225\textwidth]{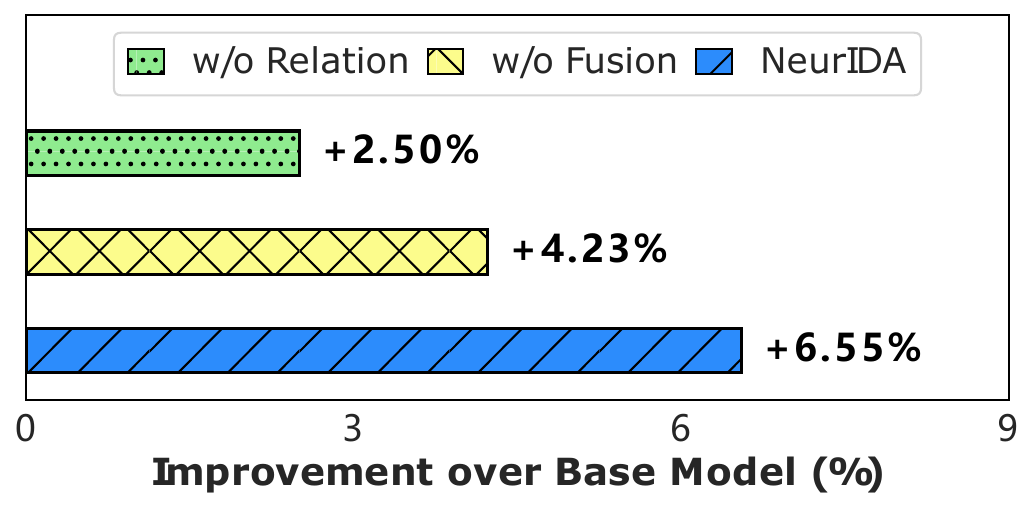}
}
\hfill
\subfloat[\textsf{HM (user-churn)}]{\includegraphics[width=0.225\textwidth]{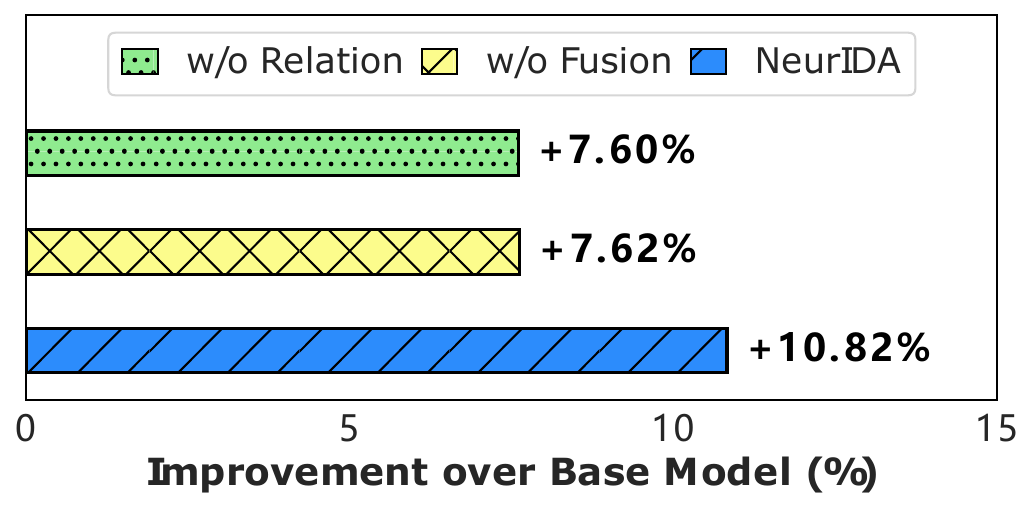}}
\\
\subfloat[\textsf{Avito (ad-ctr)}]{
\includegraphics[width=0.225\textwidth]{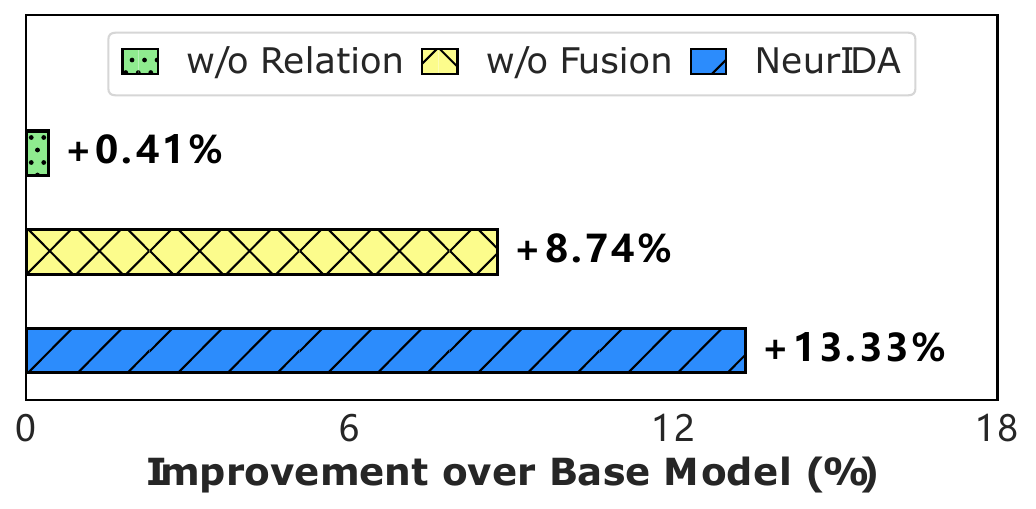}
}
\hfill
\subfloat[\textsf{HM (item-sales)}]{
\includegraphics[width=0.225\textwidth]{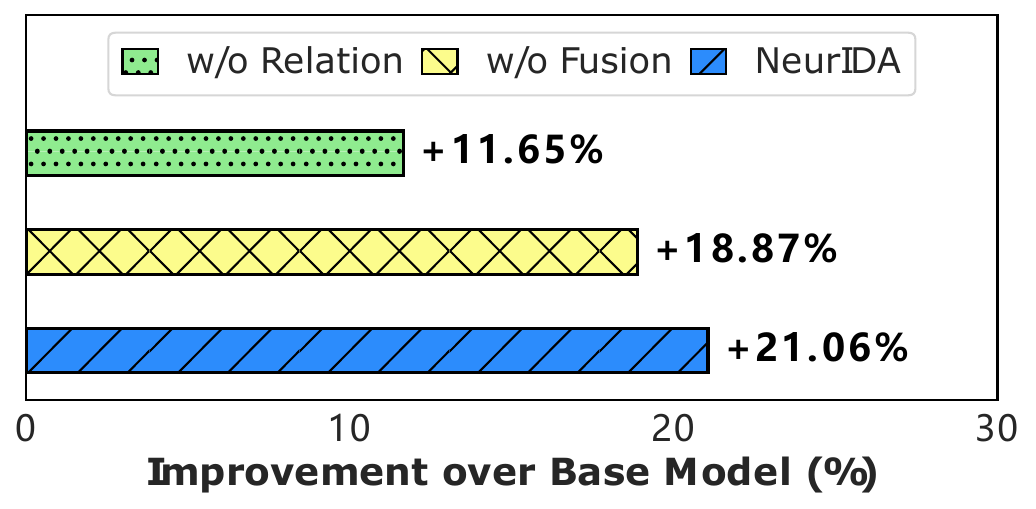}
}
\caption{Ablation study for \textit{Dynamic Relation Modeling} (w/o Relation) and \textit{Dynamic Model Fusion} (w/o Fusion).}
\label{fig:ablation}
\end{figure}

\subsubsection*{\textbf{Consistent Performance Augmentation with \oursystem{}}}
Across all prediction tasks, \oursystem{} consistently improves the performance of different base models. For classification, it brings relative gains of 4\%–12\% in \textsf{AUC-ROC}, while it reduces \textsf{MAE} by 10\%–25\% relative to the standalone base model. 
The improvements are especially pronounced for LTM approaches. For example, on the \textsf{user-repeat} task, augmenting \textsf{TP-BERTa} with \oursystem{} improves \textsf{AUC-ROC} from 0.55 to 0.77. Moreover, the enhanced model consistently outperforms strong standalone tabular models (STM) such as \textsf{TabPFN} and \textsf{TabICL}.
These gains mainly come from the dynamic modeling design in \dime. The \textit{unified tuple encoder} captures intra-table semantics and feature interactions within related tables, while the \textit{relation-aware message passing} module models inter-table dependencies across relations. The \textit{context-aware fusion} module further emphasizes task-relevant relational context for prediction. In contrast, base models without augmentation treat tuples independently and cannot exploit cross-table structure. By jointly modeling \term{intra-table semantics} and \term{inter-table dependencies}, \oursystem{} generates richer tuple representations and boosts predictive performance.

\subsubsection*{\textbf{Variability Across Model Groups}}
The performance of base models varies substantially across model groups. 
In STM, advanced tree-based models such as \textsf{CatBoost} outperform classical ML baselines like \textsf{LR} and \textsf{RF}, but do not show clear advantages over models in TRM, because high-cardinality categorical values are difficult for these models to exploit effectively without feature engineering.
By comparison, tabular foundation models, including \textsf{TabPFN} and \textsf{TabICL}, provide the strongest baselines in this group, benefiting from pre-trained priors and in-context adaptation.
TRMs show consistently strong expressiveness. Models such as \textsf{TabM} and \textsf{ARM-Net} achieve robust results, while simpler models like \textsf{DNN} perform worse in comparison. This suggests the importance of model architectural design, as attention mechanisms and expert-ensemble structures are better suited to capture complex feature interactions.
AutoML methods also achieve competitive performance by searching over large hyperparameter and architecture spaces, often approaching the best model in TRMs, although their effectiveness depends on the available computation budget.
By contrast, LTMs perform worse than other groups, revealing the limitations of language-model-based encoders for tabular analytics. One important reason is that many relational attributes, such as \texttt{UserAgentID} and \texttt{UserDeviceID} in \textsf{Avito}, are identifier-like tokens rather than semantically meaningful text, making it difficult for language models to learn informative tuple representations. Moreover, these pre-trained language models are frozen during downstream training and therefore cannot adapt to specific tasks, which further limits the capability of LTMs.
Overall, the results show that differences in architecture, feature handling, and attribute semantics strongly affect the effectiveness of base models.

\begin{figure}[t]
\centering
\subfloat[\textsf{\textsf{Avito (user-clicks)}}]{
\includegraphics[width=0.225\textwidth]{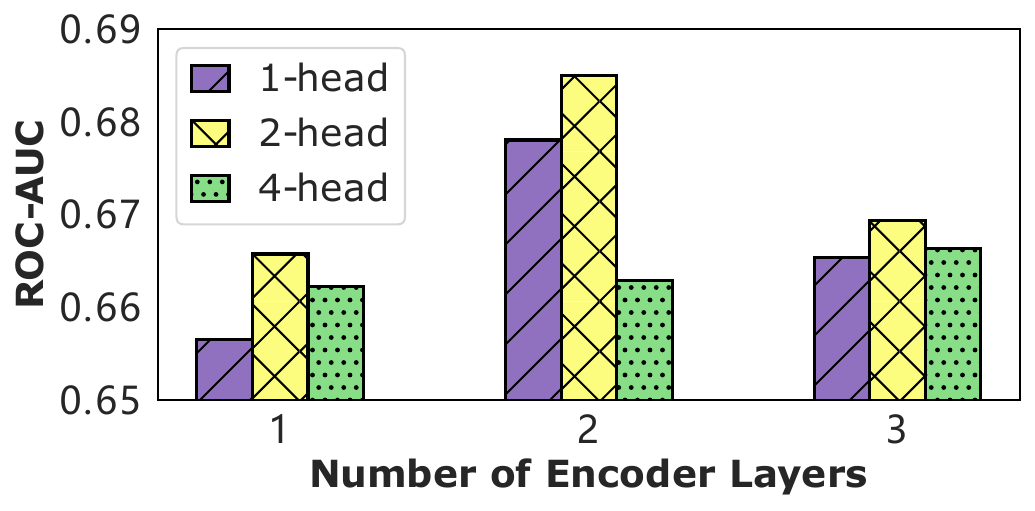}}
\subfloat[\textsf{HM (user-churn)}]{
\includegraphics[width=0.225\textwidth]{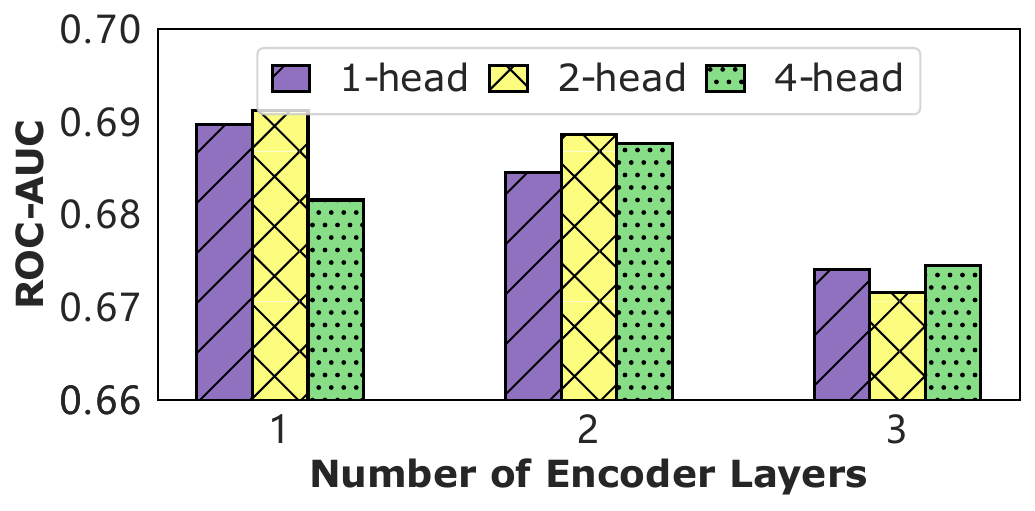}
}
\\
\subfloat[\textsf{Avito (user-clicks)}]{
\includegraphics[width=0.225\textwidth]{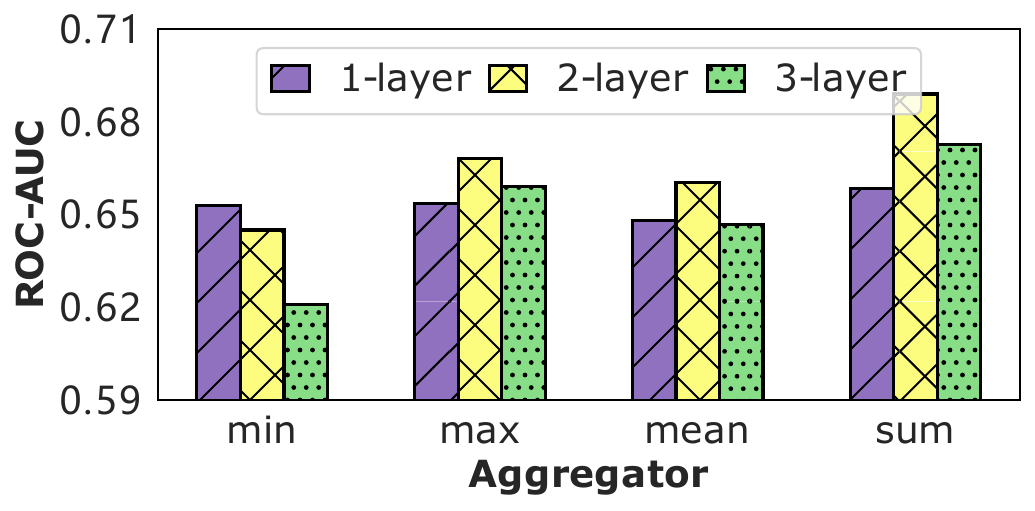}}
\subfloat[\textsf{HM (user-churn)}]{
\includegraphics[width=0.225\textwidth]{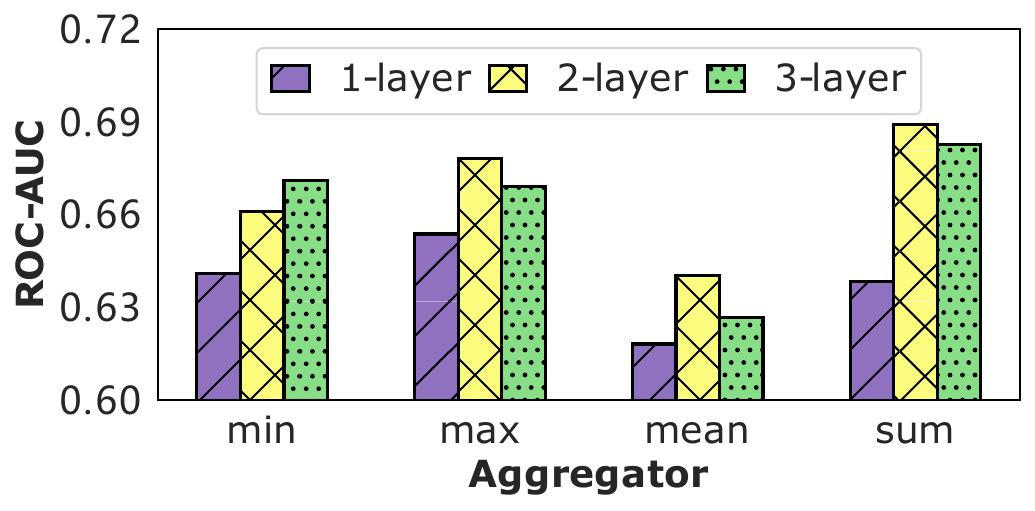}}
\caption{Effects of encoder layer $l$ and attention head $H$ (a-b),  aggregator $\textbf{agg}$ and message-passing depth $\ell$ (c-d).}
\label{fig:encoder}
\end{figure}

\subsubsection*{\textbf{Variability in Prediction Tasks}}
The augmentation brought by \oursystem{} varies across different prediction tasks. 
At a macro level, the performance gains in regression tasks are generally more substantial than in classification tasks. Regression requires precise numerical estimation and is therefore more sensitive to the quality of representation and modeling. 
The improvement achieved by \oursystem{} in regression highlights the necessity of fine-grained \term{dynamic modeling} when the prediction objective becomes challenging. 
By jointly leveraging the contextual information, \oursystem{} 
enriches tuple representations with comprehensive signals, enabling more accurate numerical prediction in regression tasks.
Additionally, the gains in classification also differ across datasets. For example, the improvement on \textsf{user-repeat} of the \textsf{Event} database is much larger than that on the \textsf{study-result} of the \textsf{Trial} database.
This discrepancy reflects how much the task depends on relational structure.
In \textsf{user-repeat}, the target table (8 features) contains limited behavior information, so most predictive signals come from related tables, making \oursystem{} particularly effective. In contrast, the \textsf{study-result} relies more on attributes already available in the target table (27 features), so the relative benefit of \oursystem{} is smaller.

\begin{figure}[t]
\centering
\subfloat[\textsf{Avito (user-clicks)}]{\includegraphics[width=0.225\textwidth]
{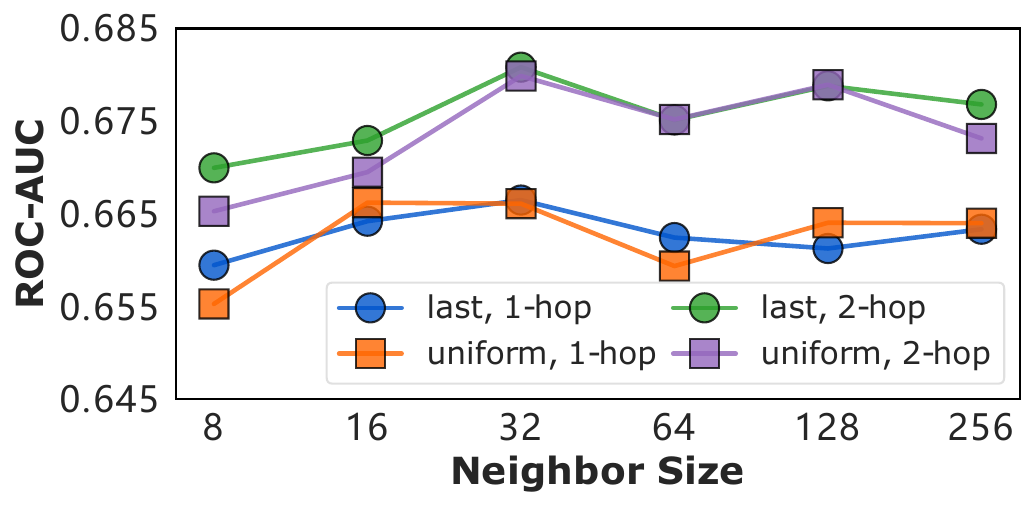}}
\hfill
\subfloat[\textsf{HM (user-churn)}]{\includegraphics[width=0.225\textwidth]
{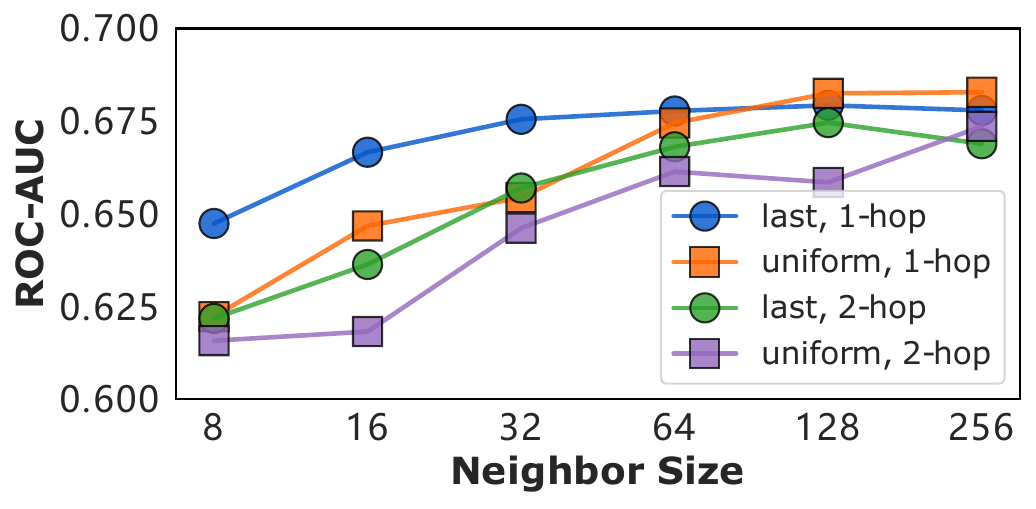}}
\\
\subfloat[\textsf{Avito (ad-ctr)}]{\includegraphics[width=0.225\textwidth]
{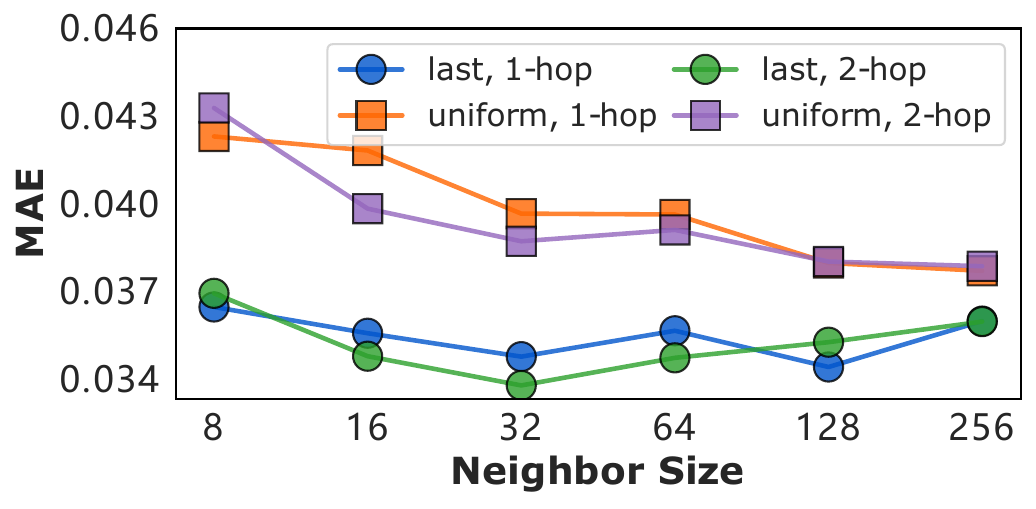}}
\hfill
\subfloat[\textsf{HM (item-sales)}]{\includegraphics[width=0.225\textwidth]
{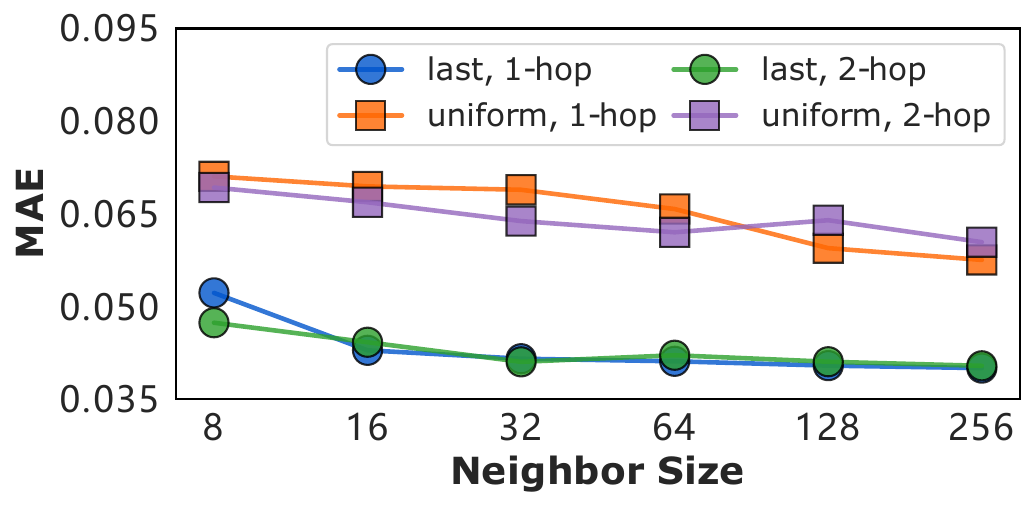}
}
\caption{Effects of neighbor sampling configuration (including neighbor size, hop number, and sampling strategy). 
}
\label{fig:neighbor-model-effectiveness}

\end{figure}


\subsection{Ablation and Parameter Analysis} \label{sec.exp.ab}
\subsubsection*{\textbf{Ablation Study}}
To analyze the contribution of different stages in \dime, we ablate the \textit{relation-aware message passing} module and the \textit{context-aware fusion} module, while retaining the \textit{unified tuple encoder} for basic tuple encoding. The two variants are denoted as \textsf{w/o Relation} and \textsf{w/o Fusion}, respectively. We evaluate their impact on the performance gain over the base model \textsf{ResNet}.
As Figure~\ref{fig:ablation} shows, removing either module degrades performance, indicating that both contribute to the effectiveness of \oursystem{}. Among them, the \textit{relation-aware message passing} module has a larger impact: without it, the performance gain drops from 6.5\% to 2.5\% on \textsf{user-clicks} and from 13.3\% to 0.4\% on \textsf{ad-ctr}. This is consistent with its role in capturing \term{inter-table dependencies} and structural information around the target table. 
We also observe that the importance of the two modules depends on schema complexity. In \textsf{HM}, which has a simple schema (3 tables and 2 relations), \textsf{w/o Relation} and \textsf{w/o Fusion} yield similar results on \textsf{user-churn}. In contrast, in \textsc{Avito} (8 tables and 11 relations), removing \textit{relation-aware message passing} leads to a much larger drop, where the improvement over the base model is reduced to only 0.4\% on \textsf{ad-ctr}.  Overall, \textit{relation-aware message passing} is the main driver for modeling complex relational signals, while \textit{context-aware fusion} provides complementary local context.

\subsubsection*{\textbf{Encoder Layer and Attention Head}}
We study the effect of encoder depth and attention head count in the \textit{unified tuple encoder} by varying the number of layers {1,2,3} and heads {1,2,4}. Results are shown in Figure~\ref{fig:encoder}. The optimal depth depends on schema complexity: a 2-layer encoder works best on \textsf{Avito} (8 tables, 11 relations), while a 1-layer encoder is sufficient on \textsf{HM} (3 tables, 2 relations). It indicates that more complex schemas benefit from higher model capacity, whereas deeper encoders may suffer from overfitting or optimization difficulty in simpler schemas. For attention heads, 2 heads give the best and most stable performance, while more heads bring little benefit but increase unnecessary complexity.

\begin{figure}[t]
\centering
\subfloat[\textsf{Avito (user-clicks)}]{
\label{analysis:user_click}
\includegraphics[width=0.225\textwidth]{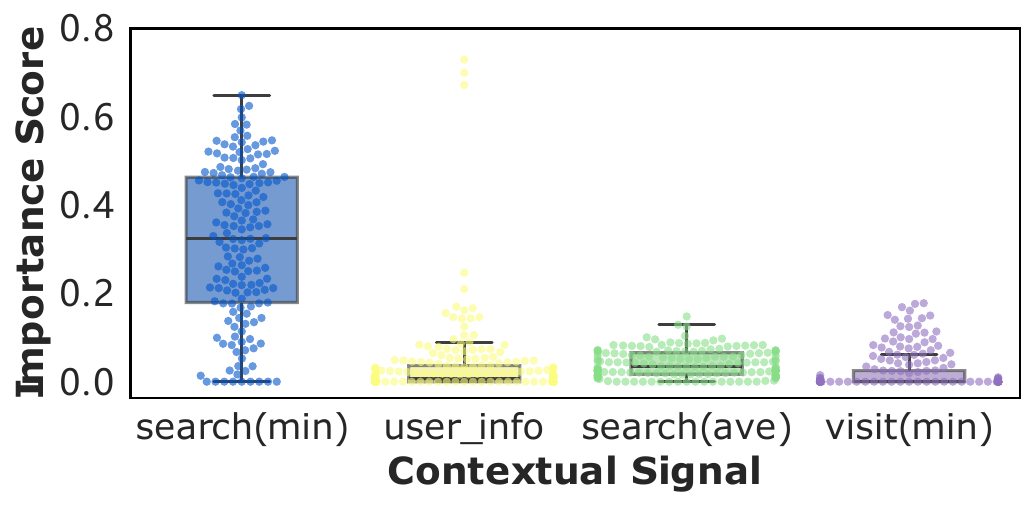}
}
\hfill
\subfloat[\textsf{HM (user-churn)}]{
\label{analysis:user_churn}
\includegraphics[width=0.225\textwidth]{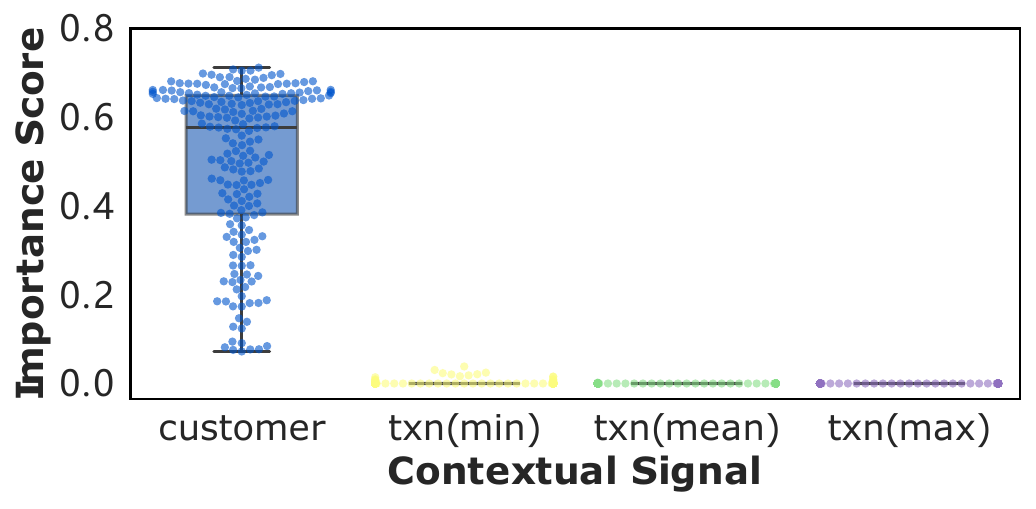}
}
\\
\subfloat[\textsf{Avito (ad-ctr)}]{
\label{analysis:ad_ctr}
\includegraphics[width=0.225\textwidth]{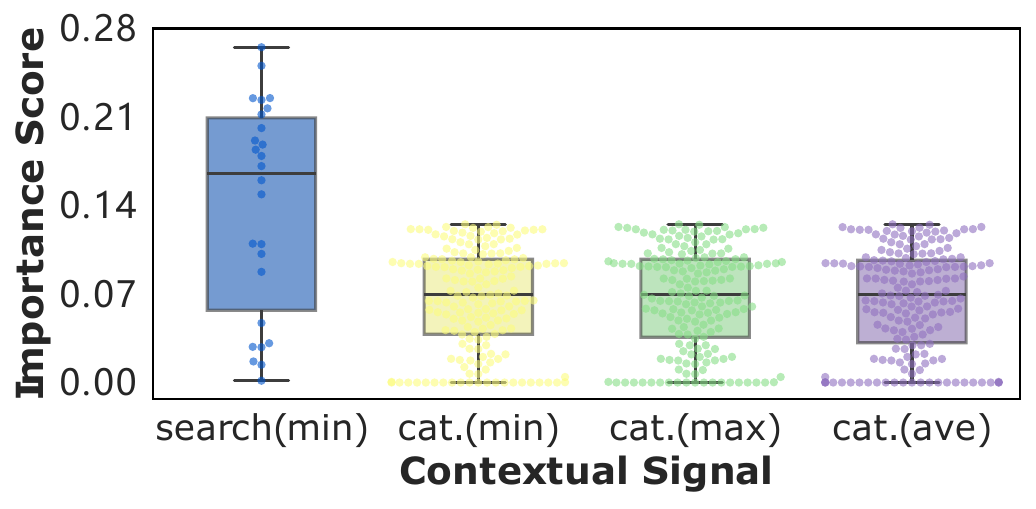}}
\hfill
\subfloat[\textsf{HM (item-sales)}]{
\label{analysis:item-sales}
\includegraphics[width=0.225\textwidth]{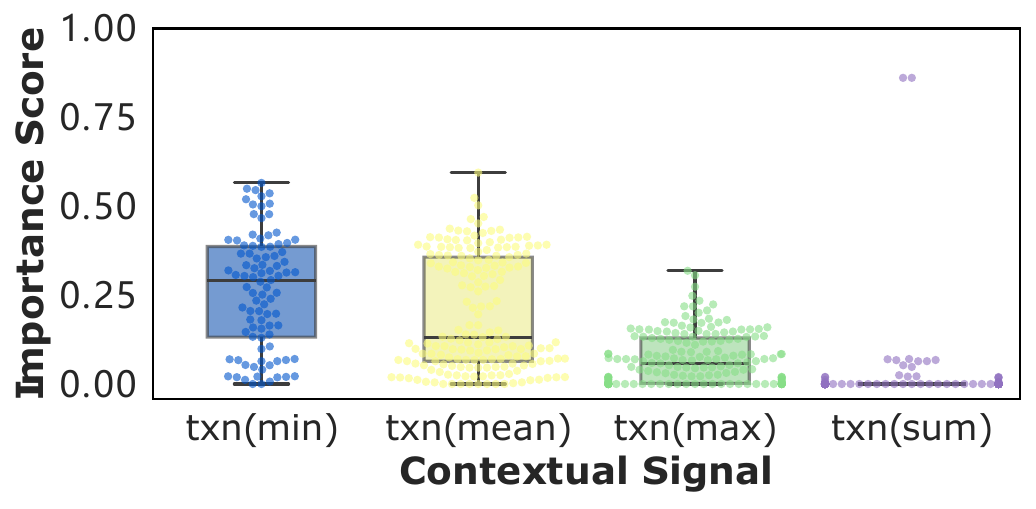}}
\caption{Interpretability visualization and analysis based on \textsf{context-aware fusion} module.}
\label{fig:neighbors}
\end{figure}

\subsubsection*{\textbf{Message-Passing Depth and Aggregator}}
We next study the depth and aggregator in the \textit{relation-aware message passing} module by varying the depth {1,2,3} and aggregators {\textsf{min}, \textsf{max}, \textsf{sum}, \textsf{mean}}.
Results in Figure~\ref{fig:encoder} show that depth 2 or 3 consistently outperforms depth 1.  It is because a larger depth expands the receptive field over the relational graph and captures multi-hop dependencies beyond directly connected neighbors. It reflects that sufficient message-passing depth is necessary to model complex relational structures and obtain expressive tuple representations. 
Among aggregators, the \textsf{sum} operation performs best overall. Unlike the \textsf{mean} operation, the default choice in most GNNs, it preserves degree-related signals that may be diluted by normalization. Such signals are particularly important in behavior-driven tasks such as \textsf{user-churn}, where the number of neighbors, e.g., user interactions, strongly reflects activity level and helps improve prediction performance.
%

\subsubsection*{\textbf{Neighborhood Sampling Configuration}}
We further study neighborhood sampling in the Dynamic Relation Modeling stage.
Since the relational graph can be large, we sample a subset of neighbors $\mathcal{N}(v)$ to efficiently capture structural information. 
We vary the \term{neighbor size}, the \term{hop number}, and the \term{sampling strategy}, including \textsf{uniform} random sampling and \textsf{latest}, which prioritizes more recent neighbors. Results are shown in Figure~\ref{fig:neighbor-model-effectiveness}.
%
Performance generally improves and then stabilizes as the \textit{neighbor size} increases, indicating that larger samples provide richer relational context but diminishing returns. Empirically, a neighbor size of 32 for the \textsf{user-clicks} and 64 for the \textsf{user-churn} is sufficient. 
The best hop number depends on schema complexity: 2-hop sampling works better on \textsf{Avito}, while 1-hop is more effective on \textsf{HM}.  This suggests that more complex schemas benefit from a larger receptive field to capture multi-hop dependencies.
%
Among sampling strategies, \textsf{latest} has the strongest advantage on regression tasks, indicating that temporally recent neighbors often provide more relevant predictive signals.

\subsubsection*{\textbf{Relation Importance}}
Finally, we examine the interpretability of the \textit{context-aware fusion} module by computing recalibrated importance scores for different contextual signals. We rank these signals by their average scores and visualize the top four in Figure~\ref{fig:neighbors}.
The learned importance aligns well with domain intuition. In \textsf{user-clicks}, \textsf{search} receives the highest score, reflecting that user search behavior is a strong indicator of future clicks, while \textsf{visit} is also useful but less influential. In \textsf{user-churn}, \textsf{user\_Info} dominates, highlighting the importance of user profiles, and in \textsf{item-sales}, \textsf{txn} is most important because historical transactions directly relate to future sales. In \textsf{ad-ctr}, \textsf{search} remains dominant, while \textsf{category} (denoted as cat in Figure~\ref{analysis:ad_ctr}) also leads to click tendencies. For example, ads in electronics and real estate often attract different levels of user attention and browsing interest.
Overall, these results show that \oursystem{} not only improves prediction but also provides interpretable insights into which relational signals drive the decision.

\begin{figure}[t]
\centering
\includegraphics[width=0.2\textwidth]{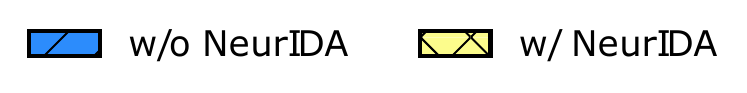}
\vspace{-0.4cm}
\\
\subfloat[\textsf{Parameter Size}]{
\label{overhead:size}
\includegraphics[width=0.225\textwidth]{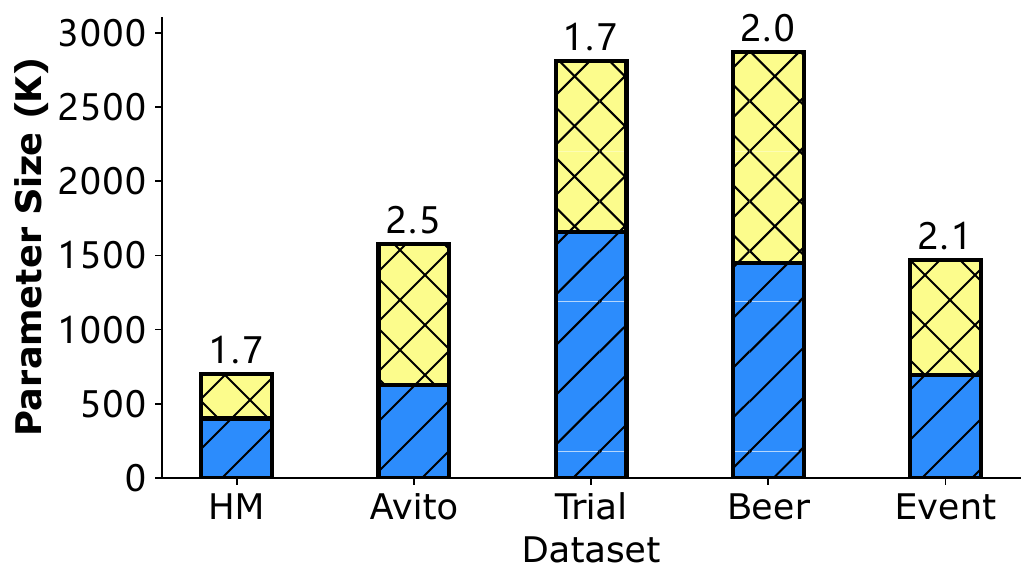}
}
\subfloat[\textsf{Latency}]{
\label{overhead:latency}
\includegraphics[width=0.225\textwidth]{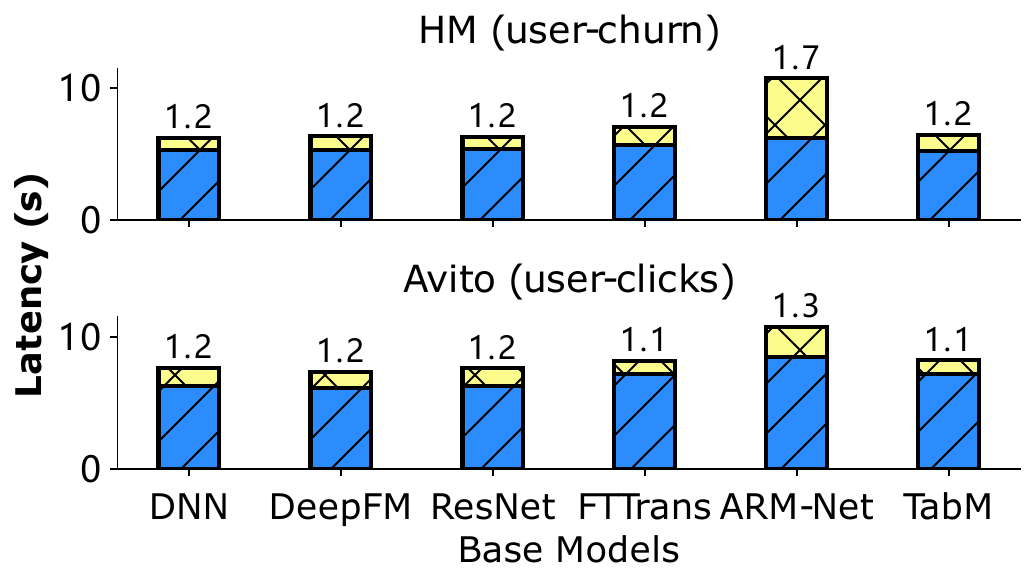}}
\vspace{-1mm}
\caption{Evaluation of \oursystem{} on computation cost. Numbers on each bar denotes the relative overhead.}
\vspace{-1mm}
\label{fig:overhead}
\end{figure}

\begin{figure}[t]
\centering
\subfloat[\revthree{\textsf{
Data Volume
}}]{
\label{scalabiliy:data}
\includegraphics[width=0.225\textwidth]{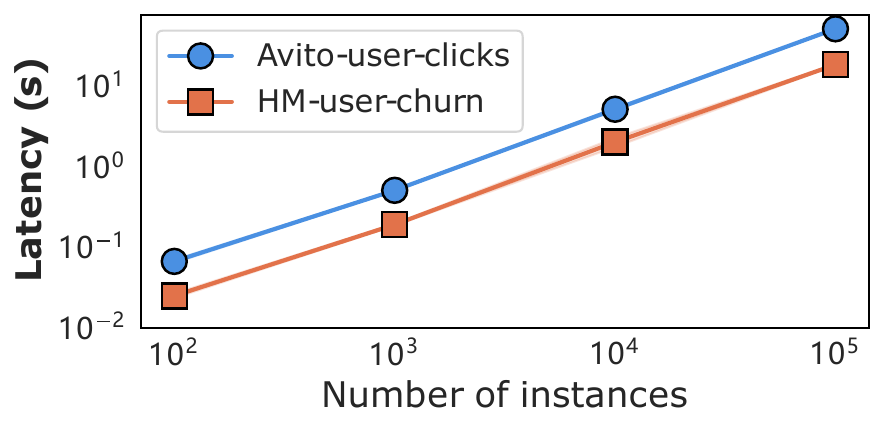}}
\subfloat[\revthree{\textsf{
Schema Complexity
}}]{
\label{scalability:schema}
\includegraphics[width=0.225\textwidth]{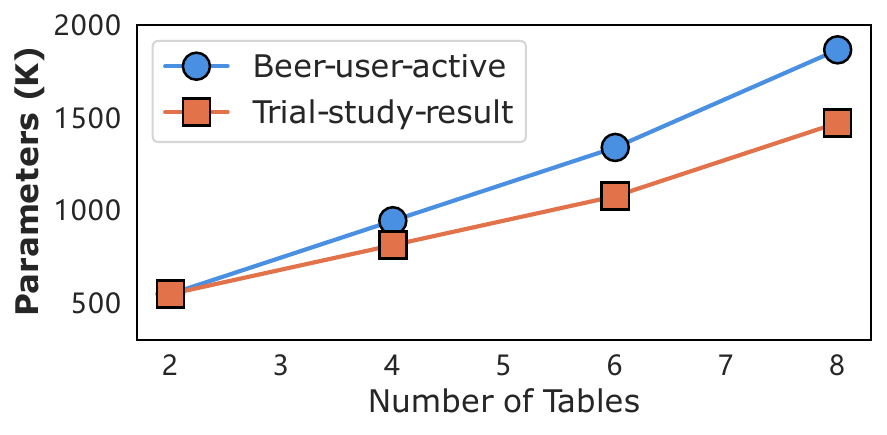}}
\caption{\revthree{
Evaluation of \oursystem{} on scalability with respect to data volume and schema complexity.
}}
\label{fig:scalability}
\end{figure}
\subsection{Cost Analysis in Different Scenarios} \label{sec.exp.cost}

\subsubsection*{\textbf{Parameter Size and Inference Latency}}
We evaluate the overhead introduced by \oursystem{}.
First, we choose \textsf{ResNet} as the \term{base model} and measure the additional parameters. As shown in Figure~\ref{overhead:size}, \oursystem{} adds between 0.7M and 3M parameters across five databases (excluding initialized embedding tables). The increase follows the complexity of the database schema: more tables and relation types require larger 
\textit{relation-aware message passing} modules. Overall, the parameter size remains within twice that of the base model.
We further measure inference latency, as illustrated in Figure~\ref{overhead:latency}. Across six different base models, \oursystem{} introduces a moderate overhead from 1.2 to 1.7 times, compared to the inference latency of base models.  Given that \oursystem{} eliminates extensive tabular preprocessing stages required by standalone base models, this additional latency remains well within a practical and acceptable range. 
Overall, \oursystem{} delivers significantly better predictive performance with only modest overhead in parameters and latency.

\subsubsection*{\revthree{\textbf{Scalability Analysis.}}}
\label{sec.exp.scalability}
\revthree{We evaluate the scalability of \oursystem{} as data volume and schema complexity increase.
For the impact of data volume, we progressively increase the size of the prediction set and measure the inference latency for the \textsf{user-clicks} task in \textsc{Avito} and the \textsf{user-churn} task in \textsc{HM}. As shown in Figure~\ref{scalabiliy:data}, the inference latency scales near-linearly with the number of prediction samples, indicating efficient and predictable scaling with respect to data volume.
To evaluate the impact of schema complexity, we incrementally expand the set of involved tables starting from the target table specified in the \emph{task profile}, and then measure the parameter size of the constructed models for \textsc{Trial} and \textsc{Beer}.
Figure~\ref{scalability:schema} shows that the model size increases approximately linearly with the number of tables.
The difference in growth slopes is due to their schema complexity, as \textsc{Beer} contains more inter-table relations, therefore requiring additional parameters in the \textit{relation-aware message passing} module.
These results demonstrate that \oursystem{} can effectively scale to larger prediction workloads and more complex schemas without sacrificing practical efficiency.
}

\begin{figure}[t]
\centering
\subfloat[\revthree{\textsf{Prediction Performance}}]{
\label{temporal:effect}
\includegraphics[width=0.225\textwidth]
{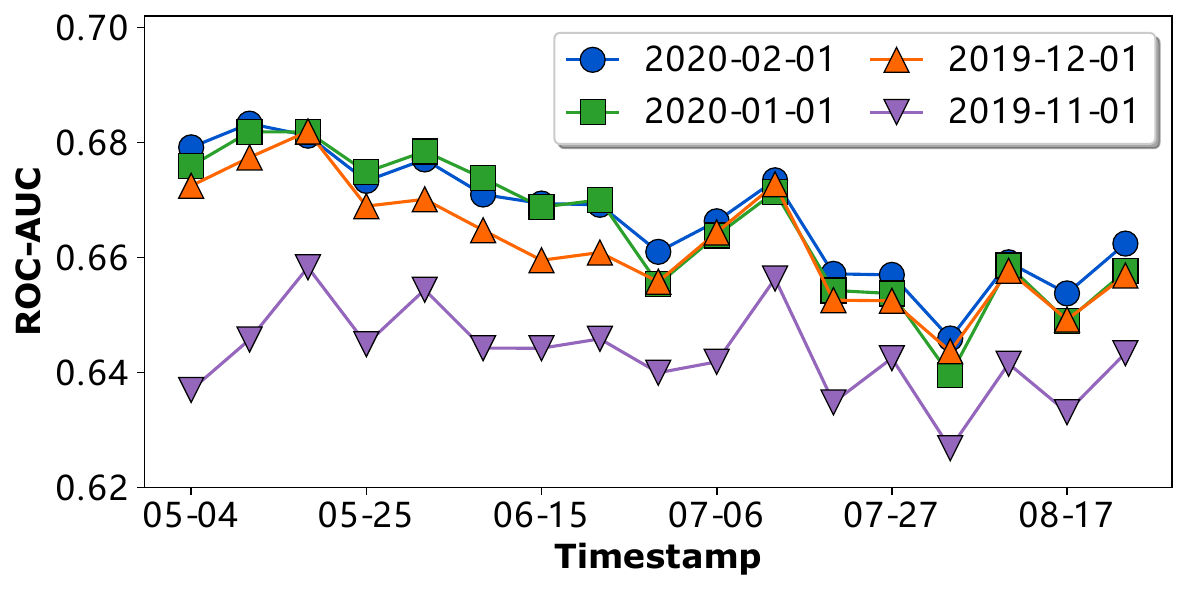}
}
\hfill
\subfloat[\revthree{\textsf{Inserted Tuples}}]{
\label{temporal:data}
\includegraphics[width=0.225\textwidth]{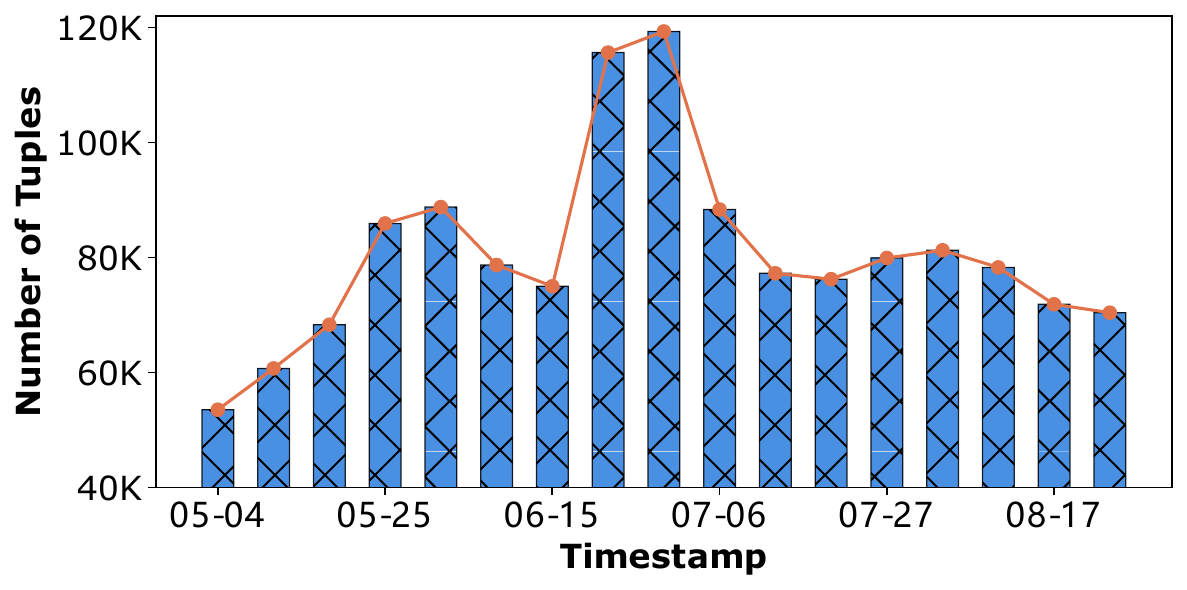}}
\caption{\revthree{
Evaluation of \oursystem{} on evolving data for the \textsf{user-churn} task on the \textsc{HM} database.
}}
\vspace{-1mm}
\label{fig:temporal}
\end{figure}

\begin{figure}[t]
\subfloat[\revthree{\textsf{Classification}}]{
\includegraphics[width=0.225\textwidth]{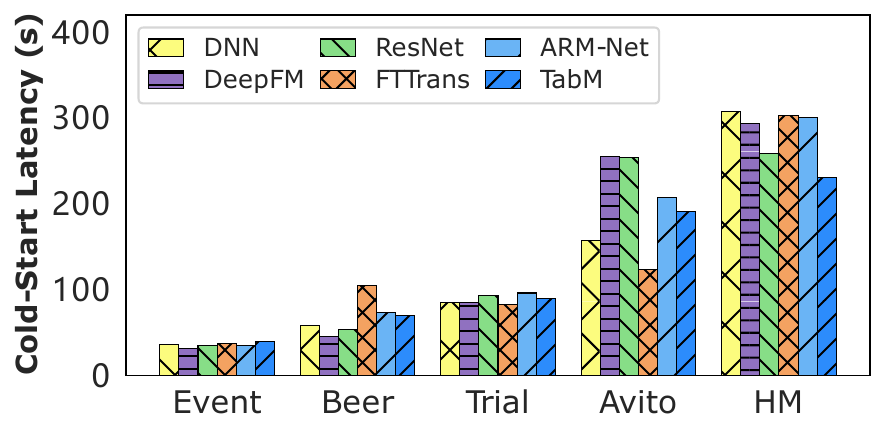}
}
\subfloat[\revthree{\textsf{Regression}}]{
\label{data_parser_level_B}
\includegraphics[width=0.225\textwidth]{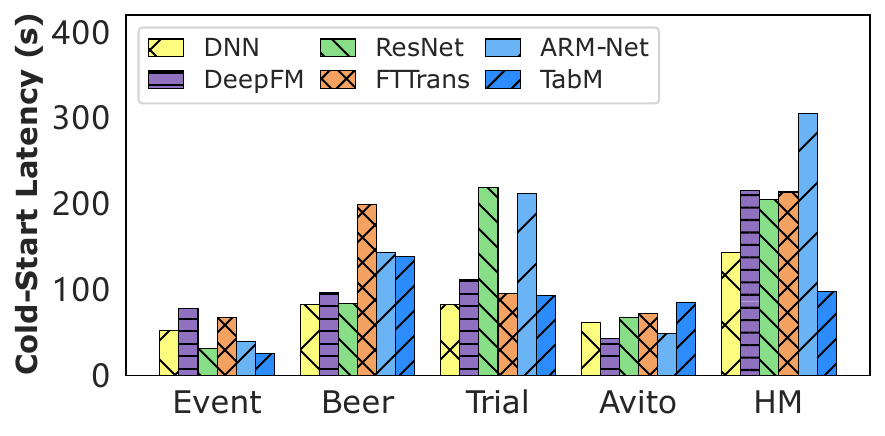}}
\caption{\revthree{
Evaluation of \oursystem{} on cold-start latency across different base models and databases. 
}}
\label{fig:cold_start}
\end{figure}

\subsubsection*{\revthree{\textbf{Evolving Data and Cold-Start Problem}}}

We study the impact of evolving data and the cold-start latency of \oursystem{}.
For evolving data, we chronologically partition each database into time windows, treat newly inserted tuples as incremental data, and evaluate prediction performance over time.
We also vary the training-data end timestamp to examine the effect of data freshness. As shown in Figure~\ref{temporal:effect}, different end timestamps provide different amounts of historical data for model construction. The results from May to August show that \oursystem{} effectively captures general behavioral patterns when sufficient historical data is available, while maintaining stable performance as data evolves.

For cold-start problem, we consider a new analytical query for which no task-specific module has been initialized.
In this setting, \oursystem{} constructs the model by selecting the base model and fine-tuning task-specific modules in \dime.
We measure this latency across all databases and base models.
As shown in Figure~\ref{fig:cold_start}, latency varies substantially across databases, mainly due to differences in data volume reported in Table~\ref{tab:data}, whereas the base model choice has limited impact.
Moreover, early stopping terminates fine-tuning once convergence is reached, keeping latency practical.
Overall, this analysis confirms that \oursystem{} remains effective under evolving data while maintaining acceptable cold-start latency in practice.

\subsection{Evaluation of Agent Components} \label{sec.exp.agent}
In this subsection, we evaluate nine LLMs from four distinct providers as backends for the agent module in \oursystem{} to assess the robustness and compatibility of the system design.

\subsubsection*{\textbf{\revone{Evaluation of Task Parser}}}

The evaluation set is built from the task profiles in Table~\ref{tab:data}. 
Positive samples are generated by varying key attributes and prompting LLMs to produce semantically equivalent natural language queries (NLQs), while negative samples cover representative failure cases, such as ambiguous targets, missing prediction objectives, and schema inconsistencies.
The benchmark contains 175 samples, including 120 positive and 55 negative instances, and is evaluated using \textit{accuracy}.
As shown in Figure~\ref{exp:task_parser}, most frontier models achieve over 94\% accuracy, indicating that \texttt{Task Parser} is reliable under schema-constrained prompting.
Larger models, such as GPT-4.1, generally outperform lightweight models, such as GPT-4o, suggesting that the task benefits from stronger reasoning.
Overall, these results confirm that the agent accurately converts natural language analytical requests into structured task profiles for downstream modeling.

\subsubsection*{\textbf{\revone{Evaluation of Data Profiler}}}

\begin{figure}[t]
\centering
\includegraphics[width=0.37\textwidth]{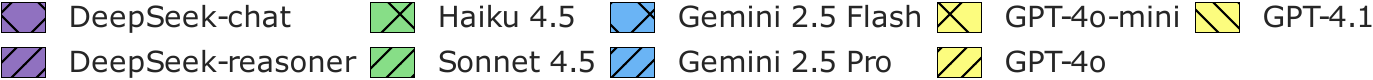}
\vspace{-0.3cm}
\\
\subfloat[\revone{\textsf{Task Parser}}]{
\label{exp:task_parser}
\includegraphics[width=0.235\textwidth]{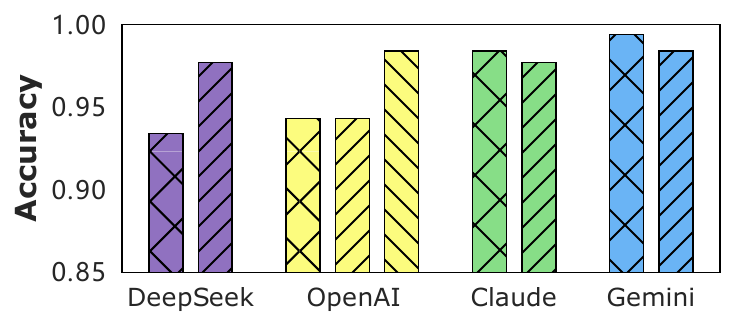}
}
\subfloat[\revone{\textsf{Data Profiler}}]{
\label{exp:data_profiler}
\includegraphics[width=0.235\textwidth]{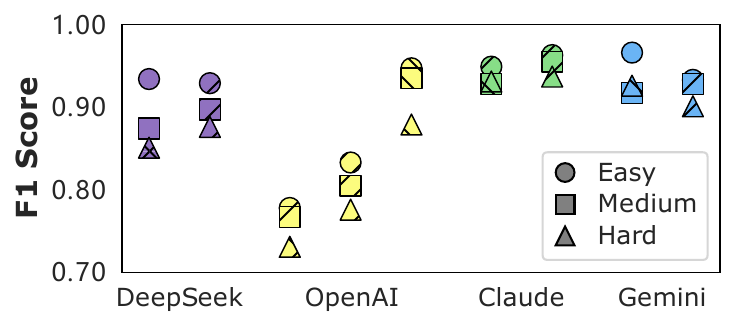}}
\caption{\revone{Performance of agent-based components 
in \oursystem{} 
across nine LLMs from four distinct providers.
}}
\label{fig:llm_agent}
\end{figure}

Starting from the five real-world databases in Table~\ref{tab:data}, we inject controlled schema noise to simulate redundant schemas under three difficulty modes: \textit{Easy}, with a few unrelated noise tables and columns; \textit{Medium}, with partially connected noise tables; and \textit{Hard}, with structurally connected noise tables and additional noise columns.
The \texttt{Data Profiler} aims to prune these irrelevant injections.
We evaluate schema selection using the column-level \textit{F1-score} of the identified task-relevant subset.
As shown in Figure~\ref{exp:data_profiler}, LLM performance degrades as schema complexity increases.
In the \textit{Easy} mode, most LLMs achieve F1 scores above 0.93.
In the \textit{Medium} mode, performance declines due to partial structural interference, while top-tier LLMs such as Claude remain robust.
In the \textit{Hard} mode, the drop becomes more noticeable, but leading models, such as Claude-Sonnet-4.5 and Gemini-2.5, still maintain strong performance.
Overall, these results show that the agent-based \texttt{Data Profiler} effectively filters irrelevant tables and columns and remains robust under increasing schema complexity.

\section{Related Work}
\label{sec:related_work}

\subsubsection*{\textbf{Tabular Data Analytics}} 
Tabular data analytics studies prediction from a single flattened table. Methods such as RF~\cite{Breiman01} and CatBoost~\cite{ProkhorenkovaGV18} remain strong baselines due to their robustness and efficiency.
Recently, deep learning models (e.g., DNN, ARM-Net~\cite{CaiZ0JOZ21}) have been used to capture nonlinear and high-order feature interactions, especially for high-cardinality categorical features.
Meanwhile, tabular foundation models such as TabPFN~\cite{tabfpn} and TabICL~\cite{tabicl} leverage pre-trained priors and in-context adaptation to generalize across tasks.
%
%
However, these approaches assume single-table inputs and require heavy preprocessing to capture the structure of relational databases. 
Our work addresses this gap with a unified augmentation framework for existing models.

\subsubsection*{\textbf{Table Discovery and Augmentation}}

Table discovery and augmentation techniques enhance prediction by identifying relevant tables and integrating auxiliary information from data lakes.
For example, DLN~\cite{datalakenavigator} detects connections across heterogeneous tables via column relevance, while RONIN~\cite{ronin} organizes large data collections hierarchically for efficient exploration.
Other work further exploits retrieved tables for augmentation: DFS~\cite{dfs} and AIDA~\cite{ARDA} automatically engineer aggregation features from joined data, while Leva~\cite{Leva} uses graph embeddings to preserve relational structure.
%
%
Unlike data lakes with uncertain relationships, relational databases provide well-defined PK-FK structures. \oursystem{} leverages this structure to deliver more accurate and reliable predictions.

\subsubsection*{\textbf{\revtwo{In-database ML system}}}

Prior work integrates ML into database systems to reduce data movement and support SQL-based analytics.
MADlib~\cite{madlib}, LEADS~\cite{leads}, and PostgresML provide in-database ML support.
Middleware 
and declarative 
frameworks, such as MindsDB~\cite{mindsdb} and SQLFlow~\cite{wang2020sqlflow}, expose external ML engines through SQL interfaces.
More integrated systems, like BigQuery ML~\cite{bigqueryml} and Raven~\cite{park2022end}, support ML execution in database-centered environments.
However, these systems are largely limited to static single-table modeling, whereas \oursystem{} directly operates on multi-table schemas and dynamically constructs models at query time.

\section{Conclusions}
\label{sec:conclusion}

We propose \oursystem{}, an end-to-end autonomous system for in-database analytics that dynamically constructs ML models tailored to various analytical tasks.
By aligning model design with query intent, relational schema, and task semantics, \oursystem{} overcomes the low \revthree{transferability} of static ML models. 
At its core is dynamic in-database modeling, which constructs models on-the-fly from a shared base model, adapting to diverse prediction objectives without manual pipeline redesign.
Additionally, \oursystem{} includes LLM-based interfaces for natural language queries and result interpretation. These integrations enable efficient, adaptive, and user-friendly analytics directly within relational databases, paving the way for more intelligent and accessible AI-powered database systems.

\begin{acks}
We dedicate this work to the memory of Professor Beng Chin Ooi, whose guidance, vision, and generosity have profoundly influenced this research. His legacy will continue to inspire us.

This research is supported by the National Research Foundation, Singapore and Infocomm Media Development Authority under its Trust Tech Funding Initiative, and the National Natural Science Foundation of China (Grant No. 624B2027). Any opinions, findings, or conclusions expressed herein are those of the author(s) and do not necessarily reflect the views of the National Research Foundation, Singapore and Infocomm Media Development Authority.
\end{acks}


\bibliographystyle{ACM-Reference-Format}
\bibliography{ref}

\end{document}